\documentclass{aa}
\usepackage{amsmath}
\usepackage{amssymb} %\gtrsim, real numbers R
\usepackage[english]{babel}
\usepackage[separate-uncertainty=true]{siunitx}
\usepackage{graphicx}
\graphicspath{{Images/}}
\usepackage{hyperref}
\usepackage{natbib}
\usepackage{booktabs} %midrule in tables
\usepackage{physics} %derivative in math mode
\usepackage{mathtools} %define in math mode
\usepackage[font=small]{caption} %linebreak in caption
\usepackage{longtable} %long table
\usepackage{bm} %bold vectors
\usepackage[title]{appendix}
\usepackage[separate-uncertainty=true]{siunitx}
\usepackage{lipsum}
\hypersetup{colorlinks=true,allcolors=[rgb]{0,0,0.8}}

\newcommand{\esass}{\texttt{eSASS}\xspace}

\newcommand{\tim}[1]{{{#1}}}

\defcitealias{Allen_2011}{Allen, Evrard, \& Mantz 2011}
\defcitealias{nfw}{Navarro, Frenk, \& White 1996}
\begin{document}

\title{The SRG/eROSITA All-Sky Survey}

\subtitle{Weak-Lensing of eRASS1 Galaxy Clusters in KiDS-1000 and Consistency Checks with DES Y3 \& HSC-Y3}

\author{F.~Kleinebreil\inst{1,2}\and S.~Grandis\inst{1}\and T.~Schrabback\inst{1,2}\and V.~Ghirardini\inst{3}\and I.~Chiu\inst{4}\and A.~Liu\inst{3}\and M.~Kluge\inst{3}\and T.~H.~Reiprich\inst{2}\and E.~Artis\inst{3} \and E.~Bahar\inst{3} \and F.~Balzer\inst{3} \and E.~Bulbul\inst{3} \and N.~Clerc\inst{5} \and J.~Comparat\inst{3} \and C.~Garrel\inst{3} \and D.~Gruen\inst{6} \and X.~Li\inst{7} \and H.~Miyatake\inst{8,9,10} \and S.~Miyazaki\inst{11} \and M.~E.~Ramos-Ceja\inst{3} \and J.~Sanders\inst{3} \and R.~Seppi\inst{3,12}\and N.~Okabe\inst{13} \and X.~Zhang\inst{3}
}

\institute{Universit\"at Innsbruck, Institut für Astro- und Teilchenphysik, Technikerstr. 25/8, 6020 Innsbruck, Austria\\
$^*$\email{florian.kleinebreil@uibk.ac.at}
\and
Argelander-Institut für Astronomie (AIfA), Universität Bonn, Auf dem Hügel 71, 53121 Bonn, Germany
\and
Max Planck Institute for Extraterrestrial Physics, Giessenbachstr. 1, 85748 Garching, Germany
\and
Department of Physics, National Cheng Kung University, 70101 Tainan, Taiwan
\and
IRAP, Université de Toulouse, CNRS, UPS, CNES, F-31028 Toulouse, France
\and
University Observatory, Faculty of Physics, LMU Munich, Scheinerstr. 1, 81679 M\"unchen, Germany
\and
McWilliams Center for Cosmology, Carnegie Mellon University, 5000 Forbes Avenue, Pittsburgh, PA 15213, USA
\and
Kobayashi-Maskawa Institute for the Origin of Particles and the Universe (KMI), Nagoya University, Nagoya, 464-8602, Japan
\and
Division of Physics and Astrophysical Science, Graduate School of Science, Nagoya University, Nagoya 464-8602, Japan
\and
Kavli Institute for the Physics and Mathematics of the Universe (WPI), The University of Tokyo Institutes for Advanced Study (UTIAS), The University of Tokyo, Chiba 277-8583, Japan
\and
National Astronomical Observatory of Japan, 2-21-1 Osawa, Mitaka, Tokyo 181-8588, Japan
\and
Department of Astronomy, University of Geneva, Ch. d’Ecogia 16, CH-1290 Versoix, Switzerland
\and
Department of Physical Science, Hiroshima University, 1-3-1 Kagamiyama, Higashi-Hiroshima, Hiroshima 739-8526, Japan
}

\date{Received---; accepted ---}

\abstract
{}
{We aim to participate in the calibration of the X-ray photon count rate to halo mass scaling relation of galaxy clusters selected in the first eROSITA All-Sky Survey on the Western Galactic Hemisphere (eRASS1) using KiDS-1000 weak-lensing data. We therefore measure the radial shear profiles around eRASS1 galaxy clusters using background galaxies in KiDS-1000, as well as the cluster member contamination. Furthermore we provide consistency checks with the other stage-III weak-lensing surveys who take part in the eRASS1 mass calibration, DES Y3 and HSC-Y3, as KiDS-1000 has overlap with both.}
{We determine the cluster member contamination of eRASS1 clusters present in KiDS-1000 based on background galaxy number density profiles, where we account for the optical obscuration caused by cluster galaxies. The extracted shear profiles, together with the result of the contamination model and the lens sample selection, are then analysed through a Bayesian population model. We calibrate the WL mass bias parameter by analysing realistic synthetic shear profiles from mock cluster catalogues. Our consistency checks between KiDS-1000 and DES Y3 \& HSC-Y3 include the 
comparison of contamination-corrected density contrast profiles \tim{and amplitudes} employing the union of background sources around common clusters, as well as the individual scaling relation results.}
{We present a global contamination model for eRASS1 clusters in KiDS-1000 and the calibration results of the X-ray photon count rate to halo mass relation. The results of the WL mass bias parameter $b_\mathrm{WL}$ obtained through mock observations show that hydro-dynamical modelling uncertainties only play a sub-dominant role in KiDS-1000. The uncertainty of the multiplicative shear bias dominates the systematic error budget at low clusters redshifts while the uncertainty of our contamination model does at high ones. The cross-checks between the three WL surveys show that they are statistically consistent with each other. This enables for the first time cosmological constraints from clusters calibrated by three state-of-the-art weak-lensing surveys.}
{}


\maketitle

\section{Introduction}
The current standard model of cosmology, which describes the evolution of the Universe as a whole, i.e. its expansion history (and future) as well as the growth of its large-scale structure (LSS), predicts that structures grow from initially small density perturbations after the Big Bang. Large, cosmological simulations such as \texttt{Magneticum}\footnote{\url{http://www.magneticum.org/}} or \texttt{IllustrisTNG}\footnote{\url{https://www.tng-project.org/}} impressively display the formation of a large-scale, web-like structure (the `Cosmic Web') in the coupled matter fluid over time. In the nodes of this structure one can 
often find massive dark matter haloes hosting massive galaxy clusters.
%, which are the most massive objects in the Universe that are gravitationally bound and largely virialized in their inner regions. 
As galaxy clusters trace the underlying Dark Matter LSS of the Universe, they present a sensitive cosmological probe \citepalias[e.g.][]{Allen_2011}. To be more precise, the halo mass function (HMF), which describes the differential number density of cluster halos as a function of their masses and redshifts, is strongly dependent on the cosmological model \citep[e.g.][]{Tinker_2008, Bocquet_2020}, and can be calibrated against cosmological simulations. Thus, accurately reconstructing the mass distribution of large samples of galaxy clusters at different redshifts (and therefore look-back times) with the goal of constraining cosmological parameters is of great interest to the astrophysical community and an active field of research \citep[e.g.][Abstract]{Pratt_2019}.

%continue here
The first catalogues of galaxy clusters were made in the 20th century, based on optically selected galaxy overdensities on the sky \citep[famously][]{Abell_1958}. However, such optical surveys are inherently affected by projection effects \citep{van_Haarlem_1997}. Galaxy clusters contain a hot plasma, the intracluster medium (ICM), which enables more robust opportunities to detect and study them via multi-wavelength observations. Next to leaving a `fingerprint' in the spectrum of the cosmic microwave background (CMB) due to inverse Compton scattering \citep[Sunyaev-Zel'dovic effect, hereafter SZ effect,][]{sz_effect}, the hot electrons that are present in the ICM emit X-ray radiation predominantly via thermal Bremsstrahlung and line emission. Therefore X-ray telescopes are a crucial tool to detect and study galaxy clusters and the LSS of the Universe. The eROSITA All-Sky Survey presents the most sensitive and highest resolution X-ray all-sky survey to date \citep{Sunyaev_2021, Predehl_2021}. Its first all-sky survey, completed in 2020, detected a total of about $\SI{12000}{}$ galaxy clusters through extended X-ray emission, although we will focus on the \textit{cosmology} sample, which provides a catalogue of more than $\SI{5000}{}$ galaxy clusters in a redshift range of $0.1<z_\mathrm{cl}<0.8$ \citep{merloni2023, bulbul23}, the largest ICM-selected cluster sample to date.

Cluster observables such as the X-ray luminosity and photon count rate are shown to scale with the halo mass both in simulations \citep[e.g.][]{angulo12, lebrun14} and in observations \citep[e.g.][]{Mahdavi_2013, bulbul19, chiu22}. 
To extract tight cosmological constrains from X-ray (or SZ-) selected galaxy cluster samples, the corresponding mass-observable scaling relations need to be well calibrated.
While halo masses can be estimated from X-ray observables \citep[e.g.][]{Reiprich_2002}, one has to make assumptions about the properties of the ICM, such as hydrostatic equilibrium, which are not well justified for general cluster samples.

Weak gravitational lensing (hereafter WL) provides a way to independently constrain halo masses purely gravitationally \citep{Schneider_2006, hoekstra2013weak}. According to General Relativity, mass curves space-time. This causes overdensities such as cluster halos to differentially deflect the light rays of more distant background galaxies propagating through their gravitational potentials. As a result the images of background galaxies (also called sources in WL terminology) that we observe experience a shape distortion, which can be measured and used to constrain the gravitational mass of the foreground cluster halo. Thus, weak lensing analyses are a vital part to calibrate mass-observable scaling relations and tighten cosmological constraints from X-ray and SZ-selected cluster surveys \citep[e.g.][]{vonderlinden14, Applegate_2014, mantz15, Hoekstra_2015, Schrabback_2018b, Herbonnet_2020, Schrabback_2021, Dietrich_2018, Zohren_2022, Bocquet_2019, bocquet23,bocquet24}.

The eROSITA All-Sky survey has overlap with several ground-based wide-area weak-lensing surveys \citep[so called Stage III surveys, see][]{albrecht2006report} in the western galactic hemisphere; the Dark Energy Survey (DES), the Hyper Suprime Cam survey (HSC), and the Kilo-Degree Survey (KiDS), see Sec. \ref{sec:data}. This presents a unique opportunity to calibrate and cross-check eROSITA's cluster cosmology analysis \citep{ghirardini23} on three independent data sets. In this work we will detail our methods of measuring the shear around eRASS1 galaxy clusters in KiDS-1000 and assess the cluster member contamination (Sec. \ref{sec:measurement}). For a description of the DES shear measurement see \citet{Grandis_2023}. Additionally, KiDS has overlap with both HSC and DES, which enables us to perform consistency checks between the surveys, based on contamination-corrected density contrast profiles \tim{and amplitudes}, as well as their individual scaling relation results (Sec. \ref{sec:consistency}). Unless noted otherwise, we carry out all necessary calculations with a standard $\Lambda$CDM cosmology ($\Omega_\mathrm{M}=0.3$, $\Omega_\Lambda=0.7, H_0=70\frac{\mathrm{km}}{\mathrm{s}\,\mathrm{Mpc}}$).
\section{Data}
\label{sec:data}
\subsection{KiDS-1000}

The Kilo-Degree Survey \citep{de_Jong_2012} is a large, \tim{$\SI{1350}{deg^2}$ optical} 
%\& near-infrared 
survey conducted using the VLT Survey Telescope \citep[VST]{arnaboldi_1998}, a $\SI{2.65}{\m}$ telescope that is located at ESO's Paranal Observatory in the Atacama Desert in Chile. VST is equipped with a specialized imaging instrument, OmegaCAM, which provides a $\SI{1}{\degree}\times\SI{1}{\degree}$ field of view and consists of 32 CCD detectors with $\SI{8}{MP}$ each that yield a pixel scale of $0\farcs214/\mathrm{pixel}$ \citep{OmegaCam}, thus being seeing-limited at all times. KiDS achieves a median seeing in the $r$-band of better than $0\farcs7$. The survey is split into two patches on the sky with roughly the same area, KiDS North and KiDS South.

\begin{figure*}[ht]
    \centering
    \includegraphics[width=\linewidth]{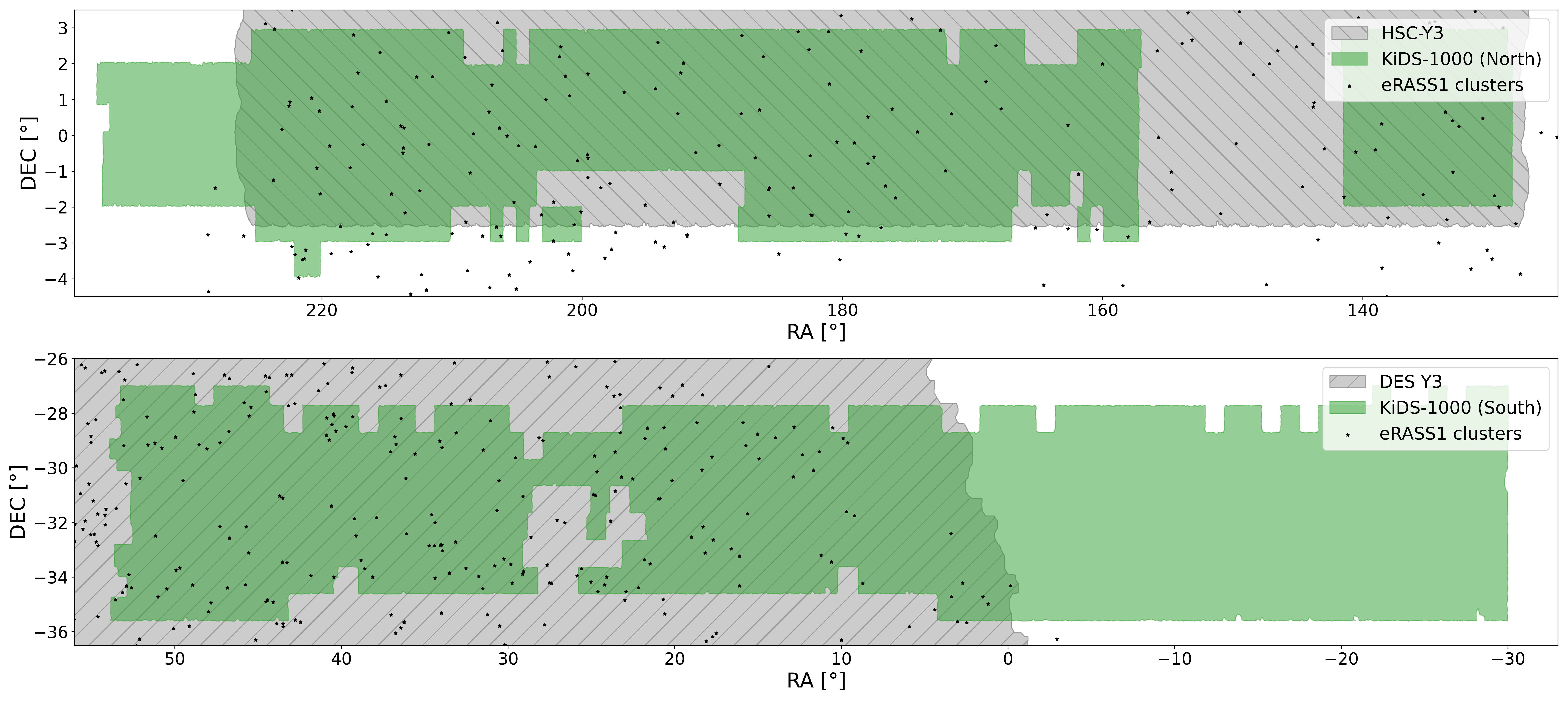}
    \caption{Footprint of KiDS-1000 North (top) \& South (bottom), as well as DES Y3 (overlap with KiDS South), HSC S19 (overlap with KiDS North), and the eRASS1 cluster sample.}
    \label{fig:footprint}
\end{figure*}

\begin{figure}[ht]
    \centering
    \includegraphics[width=\linewidth]{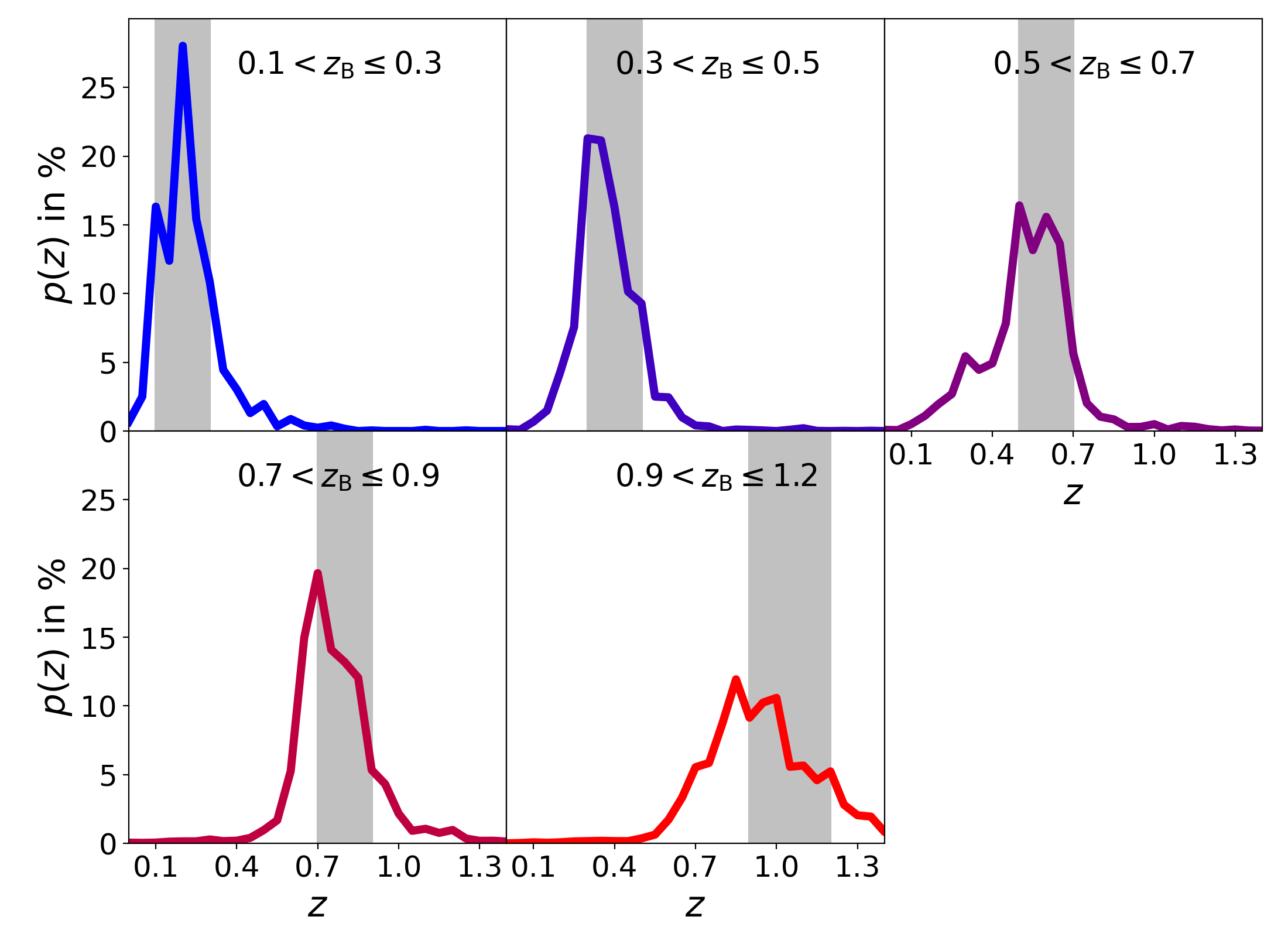}
    \caption{The five spectroscopically calibrated tomographic redshift bins of KiDS, the redshift resolution is 0.05. We color-code low-redshift tomographic bins blue, and high-redshift bins red, a convention that we perserve throughout this work.}
    \label{fig:tbins}
\end{figure}

We use the gold sample of weak lensing and photometric redshift measurements from the fourth data release of the Kilo-Degree Survey \citep{Kuijken2019, Wright_2020, Hildebrandt2021, Giblin2021}, hereafter referred to as KiDS-1000, which contains weak-lensing data for about 21 million galaxies \tim{and covers an area of $\SI{1006}{deg^2}$.}  Accurate lensing masses require accurate measurements of both galaxy shapes and their redshift distributions. For KiDS-1000, galaxy shapes were measured using the $r$-band images \citep{Giblin2021} with \textit{lens}fit \citep{Miller_2013} with calibration procedures described in \citet{Conti_2017, Kannawadi_2019}. Photometric redshifts were estimated using the KiDS $ugri$ bands plus $ZYJHK_\mathrm{s}$ NIR photometry from VISTA \citep{Sutherland_2015} as detailed in \citet{Kuijken2019}. \citet{Hildebrandt2021} describe the calibration of the source redshift distribution for KiDS-1000 galaxies split into 5 tomographic photometric redshift bins (see Fig. \ref{fig:tbins}) using self-organising maps \citep[SOMs, ][]{Wright_2020}. This approach has first been used in KIDS in the KV-450 cosmic shear analysis \citep{Hildebrandt_2020}. We show the footprint of KiDS-1000 in Fig. \ref{fig:footprint}.

\subsection{DES Y3}
The Dark Energy Survey (DES) is a large optical survey covering $\SI{5000}{deg^2}$ on the sky, having significant overlap with the KiDS-South field. In this analysis we make use of the data from the first three years of observations (hereafter DES Y3). DES is conducted using the Blanco telescope at the Cerro Tololo Inter-American Observatory (CTIO) in Chile. This $\SI{4}{\meter}$ telescope is equipped with a survey imaging instrument, DECam \citep{Flaugher_2015}, which takes data in the $grizY$-bands.

The DES Y3 weak-lensing data products provide information for roughly 100 million galaxies (after selection cuts). They include a shape catalogue \citep{Gatti_2021} generated with the \textsc{Metacalibration} pipeline \citep{huff2017metacalibration}, as well as photometric data \citep{Sevilla_Noarbe_2021}. Further information about the point-spread function modeling can be found in \citet{Jarvis_2020}, for details about image simulations see \citet{MacCrann_2021}.

Similarly to KiDS, DES makes use of a tomographic approach \citep{myles21} that was originally developed for the DES 3x2pt effort and is tightly integrated into the image simulations and calibration process.
We refer to \citet{Grandis_2023} for information about the shear measurement of eRASS1 galaxy clusters in DES Y3.

\subsection{HSC-Y3}
HSC is a deep optical imaging survey that is carried out as part of the Subaru Strategic Program using the $\SI{8.2}{\meter}$ Subaru Telescope \citep{Aihara_2017}. The Subaru Telescope is equipped with the Hyper Suprime Cam \citep{miyazaki_2015, miyazaki_2018}, a specialized imager with an FoV of $\SI{2.3}{\deg^2}$ surveying in the $grizy$-bands.

HSC aims to cover $\SI{1100}{deg^2}$ on the northern sky in its WIDE layer (there is also a DEEP and an UltraDEEP layer) and has overlap with the KiDS-North field. The $5\sigma$ limiting magnitudes of the WIDE layer are 26.5, 26.1, 25.9, 25.1, and 24.4 in the $g$-, $r$-, $i$-, $z$-, and $y$-bands, which represents the deepest imaging survey at this area to date. The HSC survey also provides great imaging quality reaching a mean seeing of $0\farcs58$ in the $i$-band, which is used for the shape measurement.

This analysis uses data products that stem from the three-year shape catalogue \citep[HSC-Y3,][]{Li_2022}, which has an effective galaxy number density of $19.9\,\mathrm{arcmin}^{-2}$. The shape measurement closely follows those methods employed in the analysis of the first-year HSC survey shape data 
%(S16A) \citepalias{mandelbaum_hscy1shape}. 
\citep[S16A,][]{mandelbaum_hscy1shape}.
We refer to 
\citet{mandelbaum_hscy1sim}
for further information about the shape calibration against simulations, and to \citet{Oguri_2017} for map level tests. Only galaxies with an $i$-band magnitude smaller than $\SI{24.5}{mag}$ are included in the shape catalogue. The HSC weak lensing analysis around eRASS1 clusters follows the same methodology as the HSC weak lensing analysis of clusters from the eROSITA Final Equatorial Depth Survey \citep[eFEDS,][]{chiu22}.

\subsection{eRASS1 galaxy cluster catalog}
The cluster sample analysed in this paper stems from the first eROSITA All-Sky Survey in the Western Galactic Hemisphere (eRASS1), which was finished on June 11, 2020. We use the cosmology sample described in \citet{bulbul23}, which consists of significantly extended X-ray sources with reliable photometric confirmation and redshift estimates in the range of $0.1<z_\mathrm{cl}<0.8$, using the DESI Legacy Survey DR10 \citep{Dey_2019}. As further detailed in \citet{kluge23}, an adaptation of the \texttt{redMaPPer} \citep{Rykoff_2014, Rozo_2015, Rykoff_2016} algorithm, a red-sequence cluster finder designed for large photometric surveys, was employed to obtain redshift and richness estimates for eRASS1 cluster candidates.

\begin{figure}[ht]
    \centering
    \includegraphics[width=\linewidth]{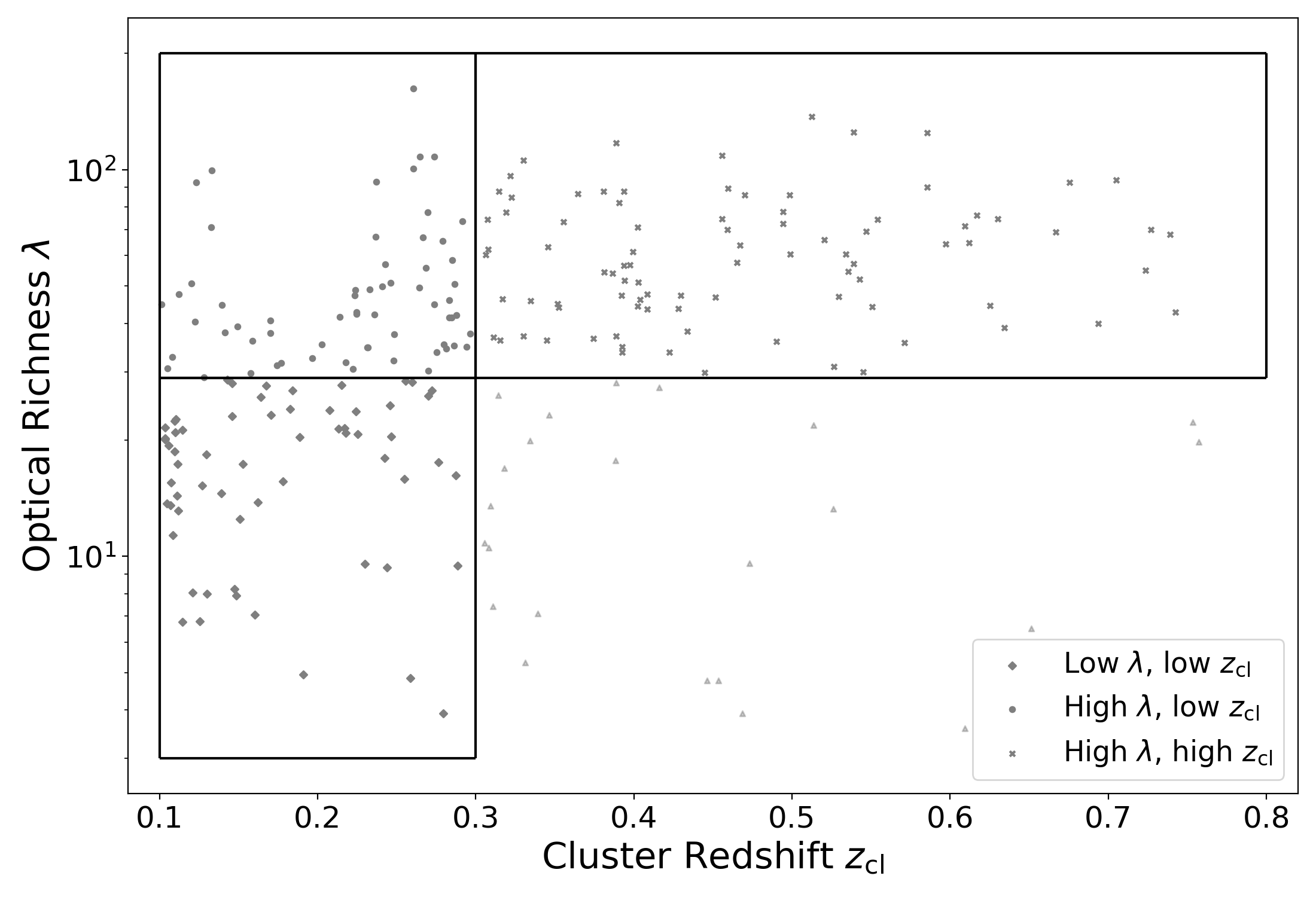}
    \caption{Richness and redshift distribution of the eRASS1 clusters located in the KiDS-1000 footprint,  including a division into subsamples for the tests presented in Section \ref{sec:kids_cont}.}
    \label{fig:clsample}
\end{figure}

In total the cosmology sample contains 5263 clusters and groups, of which 237 have overlap with KiDS-1000 (101 in KiDS-North and 136 in KiDS-South). Here, we already excluded objects that have a masked fraction larger than 50\% in the shear measurement area in KiDS-1000 from our analysis. We show the cluster redshifts and richnesses  of the KiDS-1000 eRASS1 WL sample in Fig. \ref{fig:clsample}, together with a division into three sub-samples for later testing purposes.

\section{Measurement}
\label{sec:measurement}

\subsection{Shear measurement}
\label{sec:gt_meas}
We measure the weighted reduced tangential shear of the source galaxies separately in the tomographic redshift bins. We use all tomographic bins (for a given cluster), whose lower photo-$z$ boundaries (grey bands in Fig.\thinspace\ref{fig:tbins}) are located above the cluster redshift. 
In the weak-lensing regime 
%each galaxy shape measurement represents a noisy estimate of the reduced shear at its position \citep{Bartelmann_2001}
galaxy ellipticities $\epsilon$ transform under a reduced shear $g$ as
\begin{equation}
\epsilon_\mathrm{obs}=\frac{\epsilon_\mathrm{orig}+g}{1+g^*\epsilon_\mathrm{orig}}\approx\epsilon_\mathrm{orig}+g\,,
\end{equation}
where
\begin{equation}
\label{eq:reducedshear}
g=\frac{\gamma}{1-\kappa}
\end{equation}
is related to the shear $\gamma$ and the convergence $\kappa$ \citep{Bartelmann_2001}.

Since we do not know the original ellipticities of the galaxies $\epsilon_\mathrm{orig}$ we have to average the 
%tangential 
observed 
ellipticities $\epsilon_\mathrm{obs}$ over larger background samples, and assume that their intrinsic orientations are random and not aligned with the cluster \citep[see][]{Sifon_2015}.
In this case, the average observed ellipticity
\begin{equation}
%\langle\epsilon_\mathrm{obs}\rangle\approx\langle\epsilon_\mathrm{orig}+g\rangle=\langle g\rangle\,,\quad\langle \epsilon_\mathrm{orig}\rangle=0\,,
\langle\epsilon_\mathrm{obs}\rangle=\langle\epsilon_\mathrm{orig}+g\rangle=\langle g\rangle\,,\quad\langle \epsilon_\mathrm{orig}\rangle=0\,,
\end{equation}
provides an unbiased estimate for the local reduced shear. 
%$g=\gamma/(1-\kappa)$, 
Like the ellipticity, the shear and the reduced shear can be decomposed 
%where we decompose the galaxy ellipticities 
into a tangential- and a cross-component, i.e.
\begin{equation}
\begin{split}
g_\mathrm{t} &= -g_1\cos{2\Phi}-g_2\sin{2\Phi}\\
g_\mathrm{x} &= +g_1\sin{2\Phi}-g_2\cos{2\Phi}\,,
\end{split}
\end{equation}
where $\Phi$ is the azimuthal angle of the background galaxy w.r.t. the cluster centre.

We employ 8 radial (geometric) bins with a bin width of $\SI{250}{\kilo pc}$ for our shear measurement, starting at $\SI{500}{\kilo pc}$, which leads to an outer measurement radius of $\SI{2.5}{\mega pc}$. This radial range allows us to have a good balance between measuring the signal, and being less susceptible to biases that arise from deviations of the cluster halos from perfectly centered NFW profiles
\citep*{nfw},
%\citepalias{nfw}, 
especially due to miscentering \citep[e.g.][]{Grandis_2021, sommer22, sommer2023weak}.

\begin{figure}[ht]
    \centering
    \includegraphics[width=\linewidth]{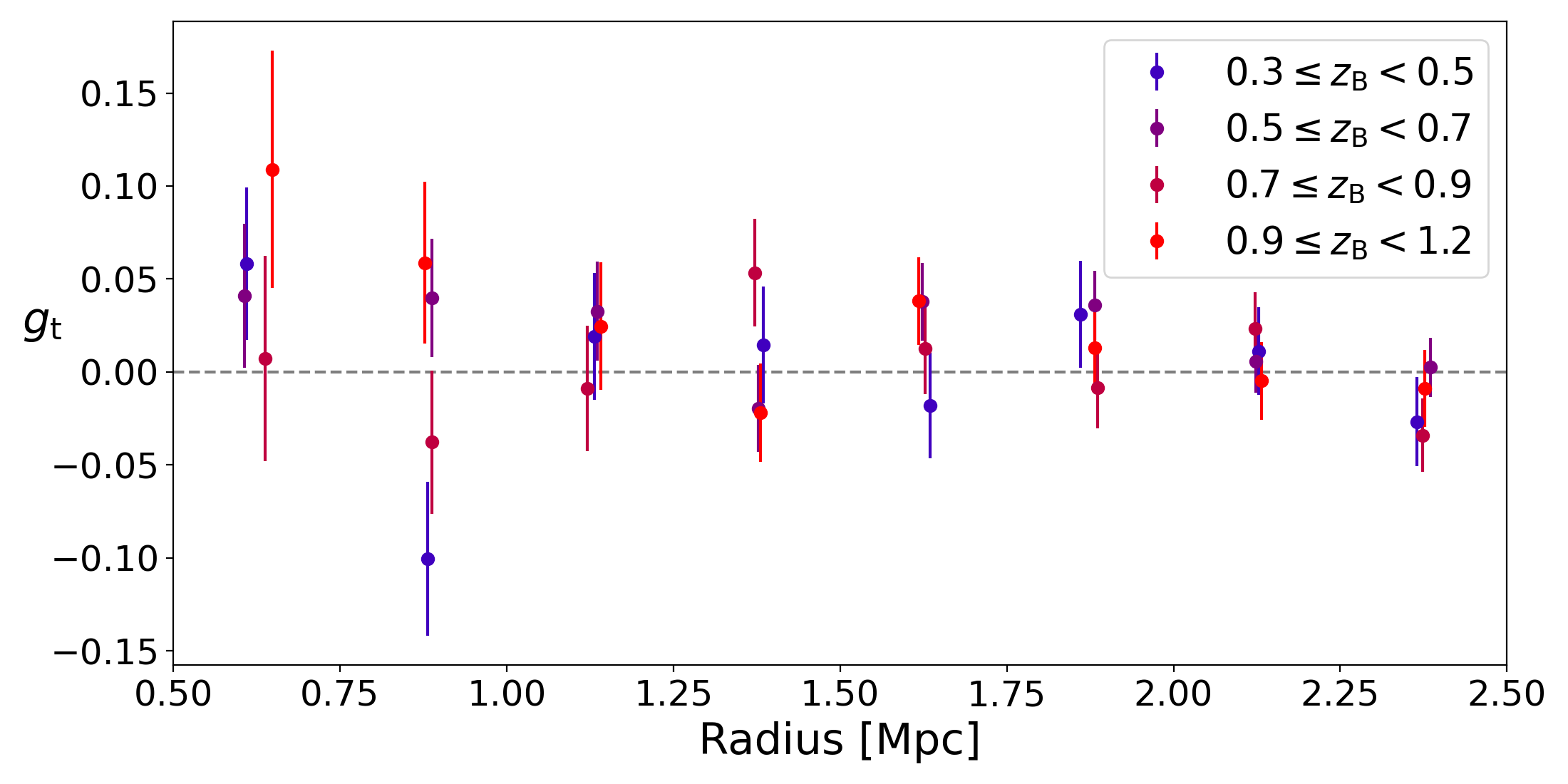}
    \caption{Exemplary radial tangential \tim{reduced} shear profile of the eRASS1 cluster em01\_019120\_020\_ML00027\_002\_c947 in KIDS-1000 with a cluster redshift $z_\mathrm{cl}=0.16$. We measure the shear separately in the tomographic background photo-$z$ bins and plot the data points at the  average projected radius of the contributing galaxies.}
    \label{fig:gt_ex}
\end{figure}

We adopt the multiplicative shear bias correction from \citet[we display the values in App. \ref{app:bwl}]{van_den_Busch_2022}, and average the tangential ellipticities in each tomographic bin $b$ and geometric annulus to obtain estimates for the tangential reduced shear
\begin{equation}
\label{eq:gt}
    g_{\mathrm{t},b}(R)=\frac{\sum_i w_i\epsilon_{\mathrm{t},i}}{\sum_i w_i\left(1+m_b\right)}\,,
\end{equation}
where $w_i$ are the individual lensing weights of the galaxies that originate from their shape measurements, $\epsilon_{\mathrm{t},i}$ are their tangential ellipticities with respect to the cluster centre, and $m_b$ is the multiplicative shear bias coefficient, which is different for each tomographic bin $b$. 
In Eq.\thinspace\ref{eq:gt} the sum runs over all galaxies $i$ which fall into redshift bin $b$ and a 250 kpc-wide annulus around a projected radius $R$. Furthermore the lensing weights in KiDS-1000 are calibrated such that
\begin{equation}
    \sigma\left(g_\mathrm{t}\right)=\frac{1}{\sqrt{\sum_i w_i}}\,.
\end{equation}
We show an exemplary tangential reduced shear profile in Fig. \ref{fig:gt_ex}.

Furthermore, visual inspection of the stacked reduced cross shear profiles of the whole cluster sample shows that they are consistent with zero and show no radial trend.

\subsection{Cluster member contamination}
\label{sec:kids_cont}
As we rely on photometric galaxy redshifts to select our background sources, cluster galaxies inevitably scatter into the selected galaxy sample that we use for our lensing measurement. These contaminating cluster galaxies suppress the shear that we measure, since they are not lensend. Depending on the quality of the photo-$z$s (and other survey properties such as depth) the strength of the contamination varies, and can become very significant, especially at small cluster-centric distances. This leads to a mass that is biased low if we do not correct for this effect.

There have been different approaches to correct for cluster member contamination in past weak-lensing analyses. For high-quality, deep data one can employ sufficient selection cuts in colour or photo-$z$ space, effectively eliminating contamination \citep{Schrabback_2018b, Schrabback_2021, chiu22}. We do not follow this approach for KiDS for two reasons. First, it would require very stringent cuts, significantly shrinking the source sample. Second, a custom source selection would require tailored redshift and shear calibration efforts, not allowing us to employ the calibration results obtained for 
%redshift distributions that have been carefully calibrated for
cosmic shear \citep{Hildebrandt2021,van_den_Busch_2022}\footnote{\tim{The stonger shears and the higher level of blending present in cluster fields may have some impact on shear calibrations. However, the analysis by \citet{hernandez-martin20} suggests that this is at a level that can be safely ignored for our present analysis given the statistical uncertainty.}}.

\citet{Gruen_2014} pioneered the decomposition of field- \& cluster galaxy populations based on their multi-band colours. This approach led to the $P(z)$ decomposition that e.g. the Dark Energy Survey prominently employs in their weak-lensing analysis \citep{Varga_2019, Grandis_2023}. This method is well applicable when analysing larger clusters samples, but it is noisy for studies analysing smaller samples or clusters individually.

Alternatively, one can calculate the number density of the selected sources and use an observed increase towards the cluster centre as a measure of the contamination \citep[e.g.][]{Hoekstra_2007, Israel_2013, Applegate_2014, Schrabback_2017, Herbonnet_2020}. This is the approach that we follow in this work for the KiDS-1000 analysis. We describe it in more detail in \citet{Kleinebreil_2023}.

\subsubsection{Radial number density profiles}
We measure the radial number density profiles of source galaxies around eRASS1 clusters in KiDS-1000 $n_\mathrm{gx}$ for each selected photometric redshift bin in $\SI{250}{\kilo pc}$ wide annuli (similar to the shear measurement), weighted with the individual galaxy lensing weights $w_i$
\begin{equation}
    n_\mathrm{gx}(R)=\frac{\sum_i w_i}{A_\mathrm{unmasked}(R)},
\end{equation}
where $A_\mathrm{unmasked}(R)$ is the unmasked geometric area of a given annulus. This incorporates the high-resolution pixel mask ($\ang{;;6}$ pixel side-length) of KiDS to account for masked areas in the shear catalogue. Furthermore we approximate the corresponding Poisson errors as
\begin{equation}
    \sigma\left[n_\mathrm{gx}(R)\right] =\frac{\sum_i w_i}{A_\mathrm{unmasked}(R)\sqrt{N_\mathrm{gx}(R)}},
\end{equation}
where $N_\mathrm{gx}$ is the number of galaxies in a given bin. We then normalise the background galaxy number density to the value that we find in the outer part of the cluster field, between $\SI{3.0}{Mpc}-\SI{3.5}{Mpc}$. We choose this radial range to measure the reference source number density, as going to larger radii can cause the cluster field of the lowest redshift clusters in the sample ($z_\mathrm{cl}=0.1$) to cover even more than four KiDS-1000 tiles. Furthermore visual inspection of the number density profiles shows that we can expect a negligible amount of contamination at these radii.

\begin{figure}[ht]
    \centering
    \includegraphics[width=\linewidth]{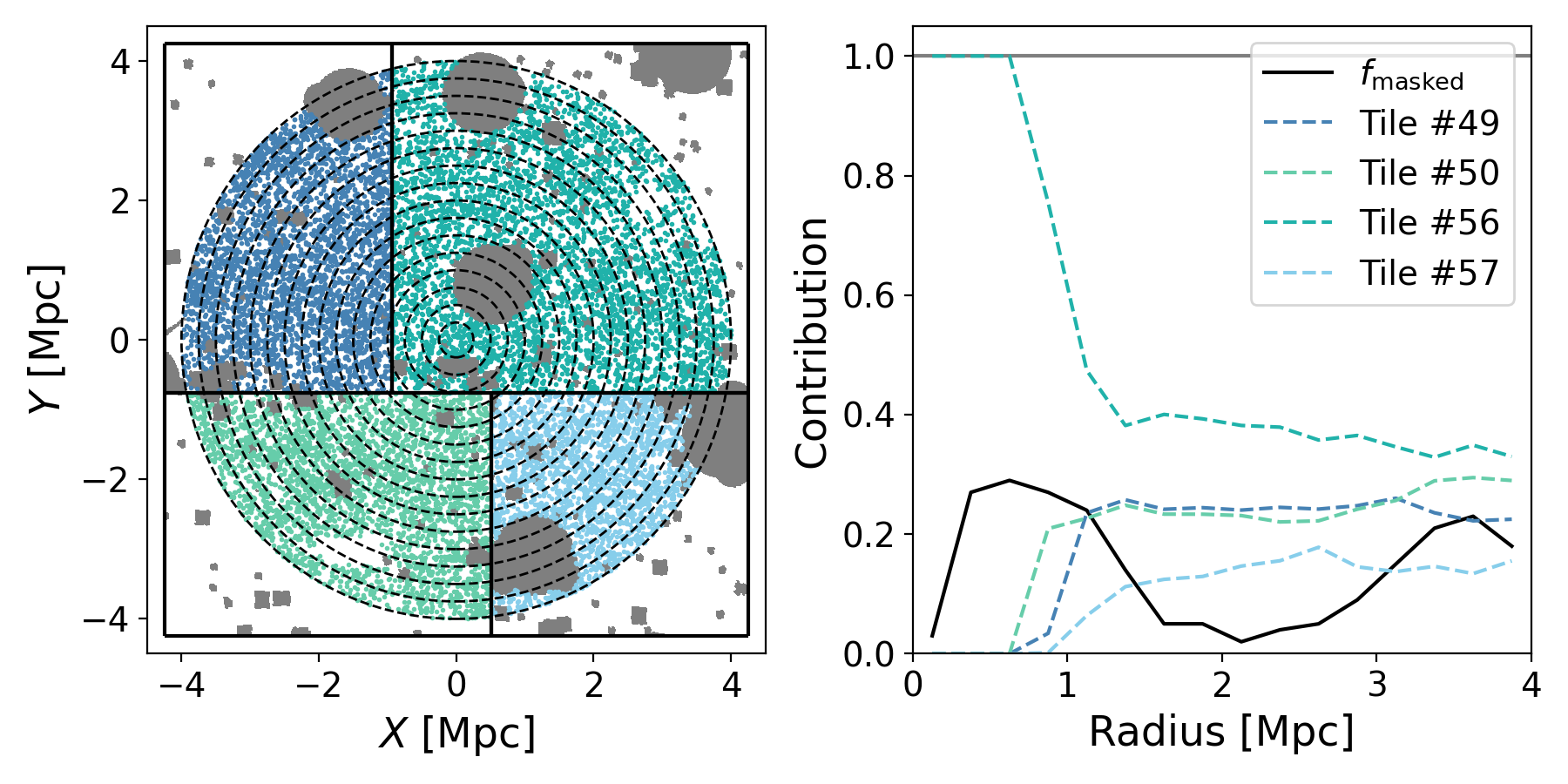}
    \caption{Left: Exemplary cluster field the of eRASS1 cluster em01\_019120\_020\_ML00027\_002\_c947 which covers four KiDS pointings. The grey areas are masked in KiDS-10000. Right: Fraction of unmasked area that each pointing contributes to a given annulus and fraction of masked pixels.}
    \label{fig:field_comp_ex}
\end{figure}
\begin{figure}[ht]
    \centering
    \includegraphics[width=\linewidth]{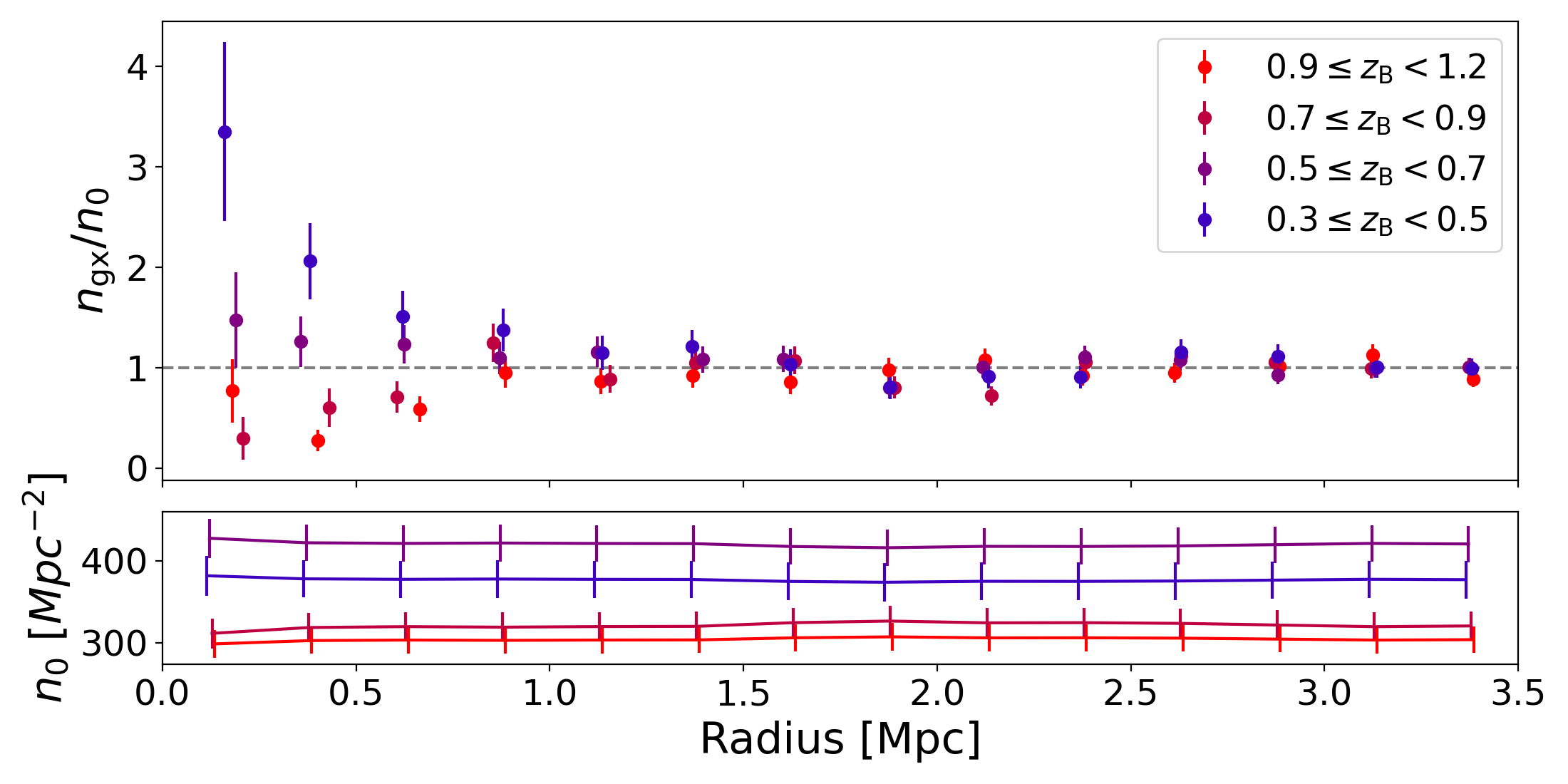}
    \caption{Top: Exemplary weighted \& normalised radial number density profile for the eRASS1 cluster em01\_046120\_020\_ML00096\_002\_c947 in KiDS-1000 with $z_\mathrm{cl}=0.25$, $\lambda=50.9$, uncorrected for obscuration. One can observe an increase in the normalised source density for the low-redshift tomographic bin towards the cluster centre due to cluster member contamination. The high-redshift tomographic bins show a decrease in source density due to obscuration from cluster galaxies. Bottom: Corresponding varying reference number density (weighted with lensing weights).}
    \label{fig:nd_ex}
\end{figure}

In case the cluster extends across multiple KiDS pointings (which can have significantly different number densities due to depth and seeing variations), we calculate the reference galaxy number density $n_0$ on a per-annulus basis. Here we measure the fraction of the unmasked area that each pointing $p$ contributes to a given annulus $A_{\mathrm{unmasked},p}(R)/A_\mathrm{unmasked}(R)$, and merge the respective reference number densities of the contributing pointings $n_{0,p}$ accordingly
\begin{equation}
    n_0(R)=\frac{\sum_p n_{0,p}\,A_{\mathrm{unmasked},p}(R)}{A_\mathrm{unmasked}(R)}\,.
\end{equation}
We show one of the cluster fields and the corresponding radial field composition in Fig. \ref{fig:field_comp_ex}, as well as the resulting radial number density profile in Fig. \ref{fig:nd_ex}. Especially the low-redshift tomographic bins show a pronounced increase in galaxy number density towards the cluster centre due to cluster member contamination, while the high-redshift tomographic bins show a decrease due to obscuration\footnote{We estimate the impact of magnification on source number densities in App.~\ref{app:magnification} and find it to be around the percent-level, which is subdominant to the impact of obscuration by cluster galaxies}.

\subsubsection{Object detection bias}
Clusters represent large optical over-densities on the sky. As a result cluster galaxies obscure the sky and impede the detection of background galaxies. 
This also affects the %radial
%can introduce an additional radial 
%profile of the selected
galaxy number density profile and can even lead to an overall 
%decreasing radial galaxy number density 
decrease
towards the cluster centre. This can especially happen in the high-redshift tomographic bins (compare Fig.\thinspace\ref{fig:nd_ex}), which contain faint galaxies and are not subject to heavy contamination in the first place.

To account for this additional bias we inject simulated KiDS-like galaxy images into the real $r$-band detection images of KiDS-1000 around each eRASS1 cluster. We use \texttt{GalSim} \citep{galsim} to generate the galaxy images as a convolution of a Sérsic profile for the galaxy itself and a Moffat profile to model the atmospheric seeing \citep{moffat}. We draw galaxies from the KiDS-1000 WL catalogue itself to attain the galaxy properties needed for the image simulation, such as half-light radius (HLR), $r$-band flux and ellipticity. An exception is the Sérsic index, which is not contained in the KiDS-1000 WL catalogue. We draw it randomly from the 2DPHOT KiDS structural parameter catalog \citep{2dphot, roy2018, amaro2021}. 
However the KiDS-1000 catalogue 
%'s galaxy population
is not complete at faint magnitudes. To simulate a complete population of galaxies based on the KiDS-1000 WL catalogue we add a selection weight based on the galaxies' $r$-band magnitude and HLR, such that we draw faint and extended galaxies (which have lower signal-to-noise ratios) more often. We obtain this selection weight by injecting $\sim\SI{2e+6}{}$ galaxies of different magnitudes and HLR's into KiDS-1000 $r$-band images and analysing their detection probability w.r.t. these quantities \citep[for more details see][]{Kleinebreil_2023}.

\begin{figure}[ht]
    \centering
    \includegraphics[width=\linewidth]{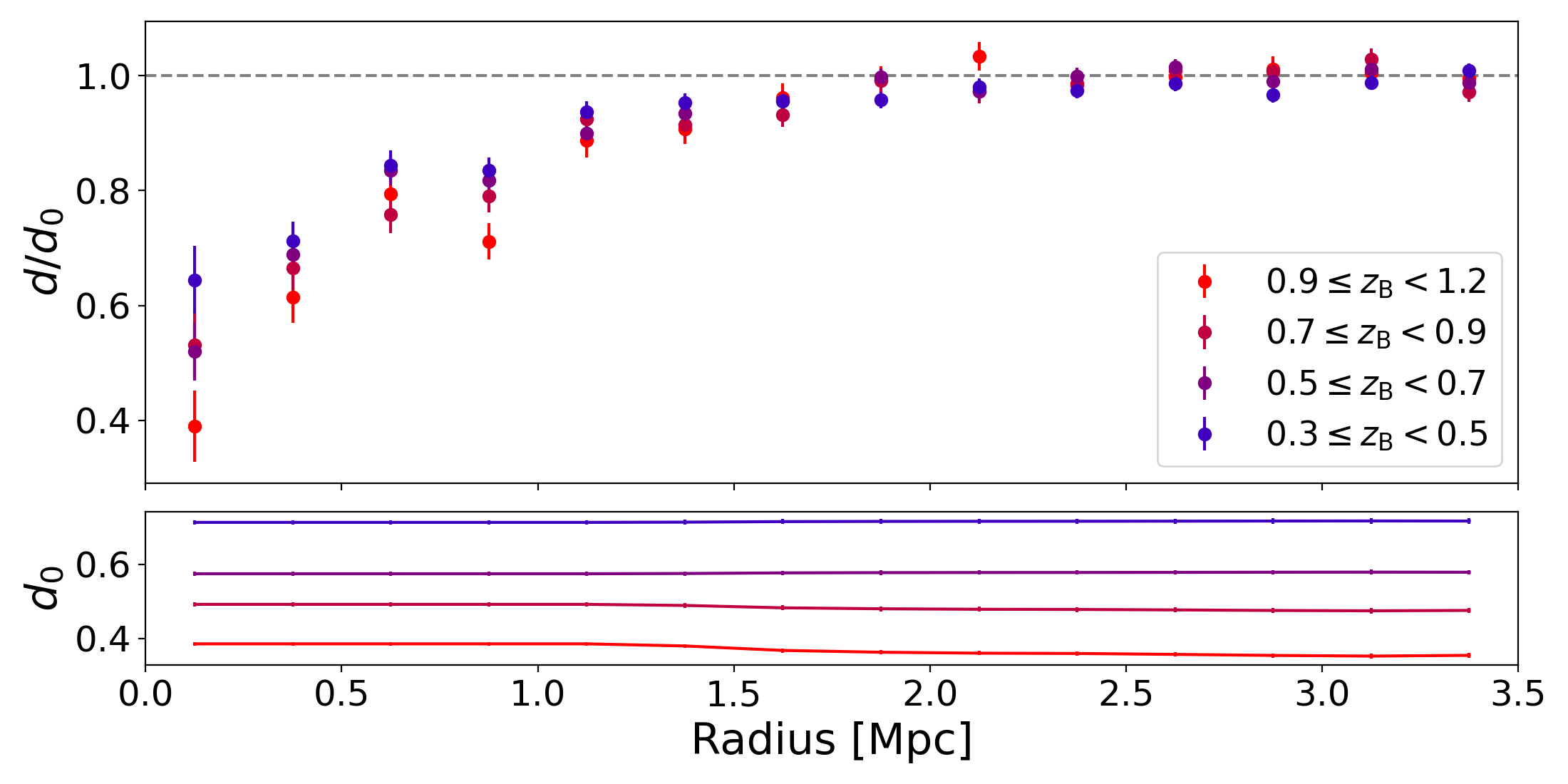}
    \caption{Top: Exemplary weighted radial detection probability profile of the eRASS1 cluster em01\_022120\_020\_ML00040\_002\_c947. Bottom: Corresponding varying reference detection probability (weighted with lensing weights).}
    \label{fig:pdet_ex}
\end{figure}

We perform several image injection runs around each cluster, adding 15\% additional galaxies in each tomographic bin in each run (relative to the overall number densities of the corresponding KiDS pointing in the different tomographic bins), until we injected at least 50,000 galaxies in each tomographic bin. We then re-run the object detection software \texttt{SExtractor} \citep{sextractor} with the settings which the KiDS-consortium used to produce the KiDS-1000 WL catalogue \citep[see][]{Kleinebreil_2023}. Since we know at which positions we injected the galaxy images (as well as their original lensing weight in the WL catalogue) we can calculate the weighted radial detection probability profile $d(R)$ for each cluster. We investigate the average lensing weight per galaxy in the KiDS-1000 WL catalog around galaxy clusters as a function of cluster-centric distance in \citet{Kleinebreil_2023}, and find no statistically significant change. We therefore argue that blending does no affect the lensing weights significantly in the context of this analysis.

For this we average the ratio of re-detected to injected galaxies at a given annulus (and tomographic bin) over all injection runs, weighted with the lensing weights. Since the positions of the injected galaxies are random the individual injection runs will inject a varying number of galaxies in a given annulus. Thus, we weight the detection ratios of the individual injection runs $d_k(R)$ at a given annulus (and tomographic bin $b$) with the number of injected galaxies $N_{\mathrm{inj},k}(R)$ of that run (into this annulus and tomographic bin) when we average over all runs $k$
\begin{align}
    d_k(R) &= \frac{\sum_{i \in \mathrm{detected}}  w_{i}}{\sum_{i\in \mathrm{injected}} w_{i}},\\
    d(R) &= \frac{\sum_k d_k(R) N_{\mathrm{inj},k}(R)}{\sum_kN_{\mathrm{inj},k}(R)},
%    w_i &= \mathrm{lensing\ weights}.\nonumber
\end{align}
where the $w_i$ are again the lensing weights.
We calculate the corresponding run-to-run variance as uncertainty
\begin{equation}
    \sigma(d) = \sqrt{\left(\frac{\sum_k N_{\mathrm{inj},k}\,d_k^2}{\sum_k N_{\mathrm{inj},k}}-d^2\right)\frac{\sum_k \left(N_{\mathrm{inj},k}^2\right)}{N_{\mathrm{inj},\mathrm{tot}}^2-\sum_k\left(N_{\mathrm{inj},k}^2\right)}}.
\end{equation}
Furthermore we normalise the weighted radial detection probability profiles analogously to the number density profiles, employing a reference detection probability $d_0(r)$ on a per-annulus basis, which we measure in the same outer radial range $\SI{3.0}{}-\SI{3.5}{\mega pc}$ for each KiDS-1000 pointing that contributes to the cluster field (see Fig. \ref{fig:pdet_ex} for an example).\\

We fit all individual detection profiles simultaneously with a global model that we base on the cluster redshifts and optical richnesses $\lambda$ according to
\begin{equation}
    \frac{d(R)}{d_0(R)}=\frac{1}{1+f(z_\text{cl})\left(\frac{\lambda}{25}\right)^{\alpha}e^{1-R/R_\mathrm{s}}},\quad R_\mathrm{s}=R_0\left(\frac{\lambda}{25}\right)^{\beta}\,,
\end{equation}
where $R_\mathrm{s}$ is a characteristic scale radius 
%(not to be confused with the scale radius of the NFW profile) 
and $\alpha$, $\beta$, and $R_0$ are parameters of the global fit. We find the following best-fit parameters: $R_0=\SI{0.734(17)}{Mpc}$, $\alpha=\SI{0.264(37)}{}$, $\beta=\SI{-0.146(27)}{}$. The negative exponent in the relation of the scale radius is a result of the degeneracy between $\alpha$ and $\beta$.

\begin{figure}[ht]
    \centering
    \includegraphics[width=\linewidth]{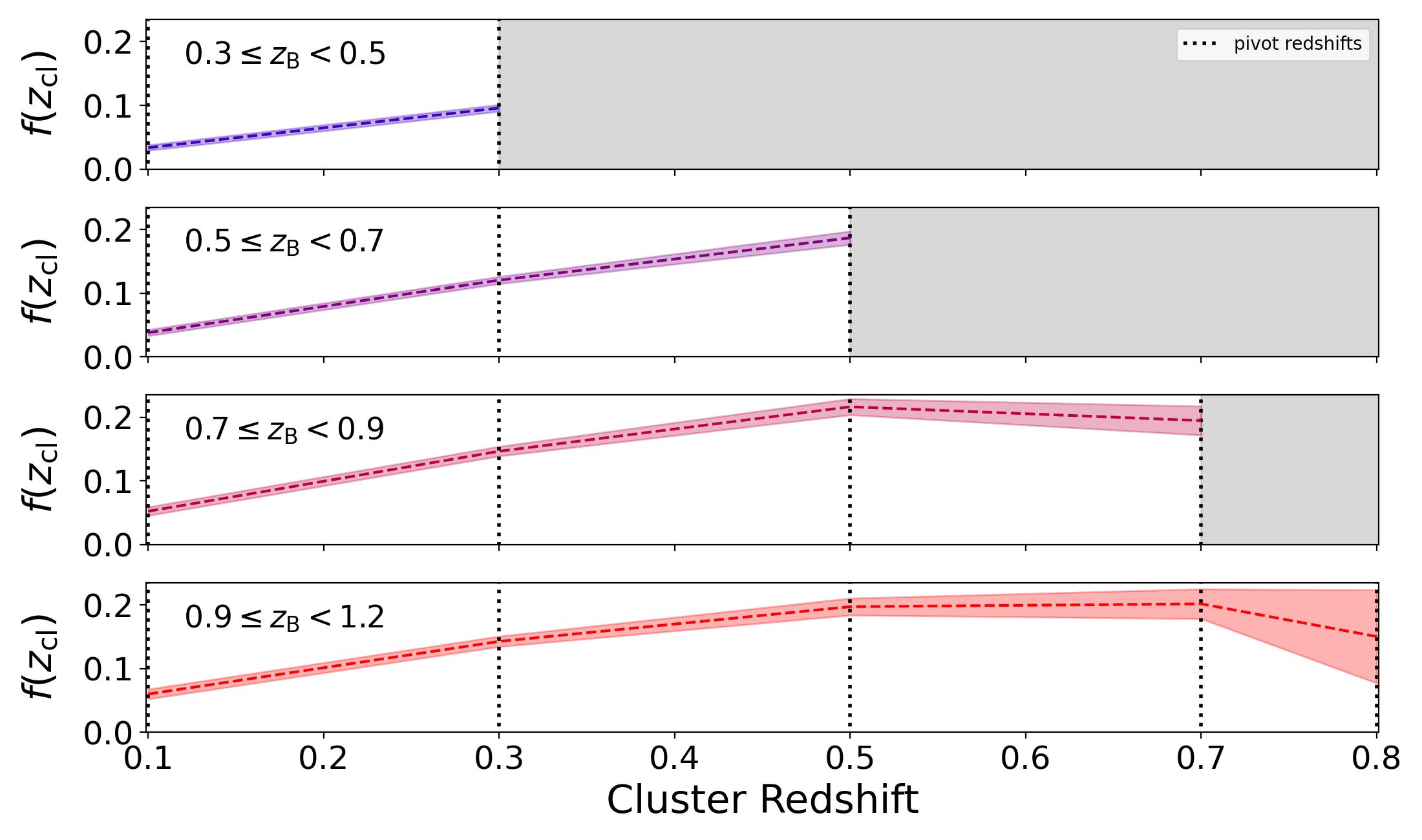}
    \caption{Best-fit parameters of the redshift evolution of the object detection bias from the global fit of all individual radial detection profiles.}
    \label{fig:d_zcl}
\end{figure}

We calculate the amplitude of the object detection bias $f(z_\mathrm{cl})$ for every tomographic background galaxy bin based on a linear interpolation between pivot points in the cluster redshift space, which also are parameters of the global fit (Fig. \ref{fig:d_zcl}). The pivot points have a redshift separation of 0.2, except for the last point in the highest redshift tomographic bin ($\Delta z=0.1$), as the eRASS1 cosmology cluster sample does not extend to redshifts above 0.8.

\begin{figure*}
    \centering
    \includegraphics[width=\textwidth]{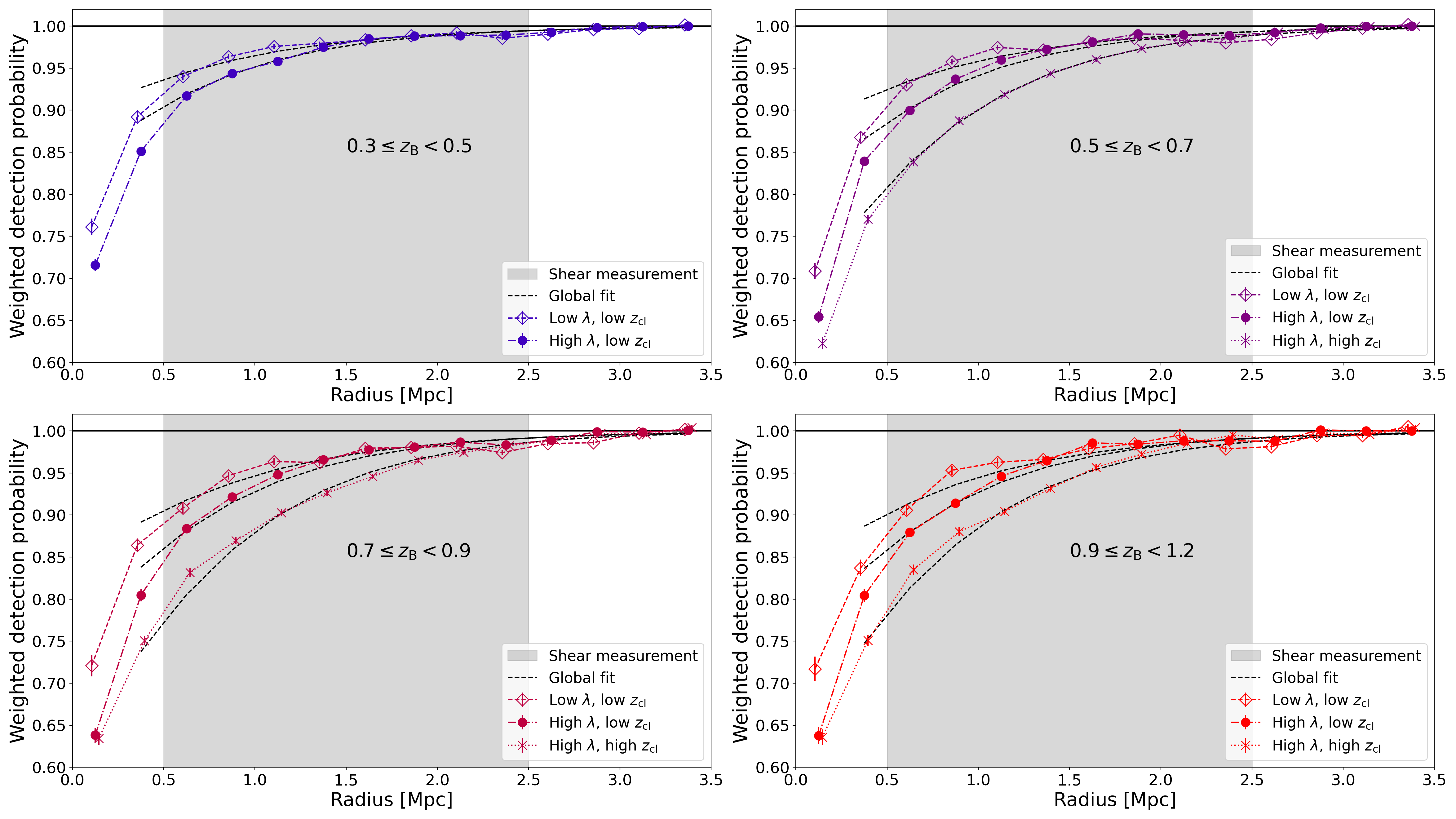}
    \caption{Measured and best-fit radial detection profiles, stacked separately for the different tomographic source bins (panels) and divided into three cluster sub-samples.}
    \label{fig:pdet_model}
\end{figure*}

To check if our global model fits the data we split the eRASS1 cluster sample in three sub-samples in cluster redshift/optical richness bins (cf. Fig. \ref{fig:clsample}). We average all measurements (clusters of the sub-sample) at each radius, and calculate the corresponding model values from the best-fit detection model based on the individual cluster redshifts and richnesses. We also average the calculated model values and show the result in Fig. \ref{fig:pdet_model}. We exclude the innermost annulus $\SI{0}{}-\SI{250}{kpc}$ (as well as the radial range of the baseline detection probability measurement $\SI{3.0}{}-\SI{3.5}{Mpc}$) from the fitting range, as the measured profiles drop steeply at the cluster centre. We argue that this is due to the brightest cluster galaxies (BCGs), which are usually located at the cluster centre and cause large object detection biases.

We observe that the strength of the object detection bias increases with cluster redshift and richness in all tomographic bins. It also increases slightly for high-redshift tomographic bins compared to low-redshift ones. At a cluster-centric distance of $\SI{1}{\mega pc}$ the weighted radial detection probability profiles typically show an effective object detection bias of 5-12\%. At $\SI{500}{\kilo pc}$ this increases to up to 20\% for the high-redshift, high-richness cluster sub-sample.

Furthermore we can confirm that the global model describes the measured detection profiles well.

\subsubsection{Contamination model}
\label{sec:cont_model}
We subsequently use the best-fit detection probability model to boost the measured radial number density profiles based on the individual cluster redshifts and richnesses
\begin{equation}
    \left(\frac{n(R)}{n_0(R)}\right)^\mathrm{corr}=\frac{n(R)}{n_0(R)}\left[1+f(z_\text{cl})\left(\frac{\lambda}{25}\right)^{\alpha}e^{1-R/R_\mathrm{s}}\right].
\end{equation}
We apply Gaussian error propagation, and in turn perform a second global fit on these boosted number density profiles to extract the final contamination model
\begin{align}
    \left(\frac{n(R)}{n_0(R)}\right)^\mathrm{model}&=\frac{1}{1-f^\prime(z_\text{cl})\left(\frac{\lambda}{25}\right)^{\alpha^\prime}e^{1-R/R_\mathrm{s}^\prime}},\\
    R_\mathrm{s}^\prime&=R_0^\prime\left(\frac{\lambda}{25}\right)^{\beta^\prime}.\nonumber
\end{align}
We find $R_0^\prime=\SI{0.69(14)}{Mpc}$, $\alpha^\prime=\SI{0.31(37)}{}$, $\beta^\prime=\SI{-0.06(24)}{}$. The uncertainties are larger than for the detection model, because we only have one realisation for each cluster, compared to several hundred image injection runs.\\

\begin{figure}[ht]
    \centering
    \includegraphics[width=\linewidth]{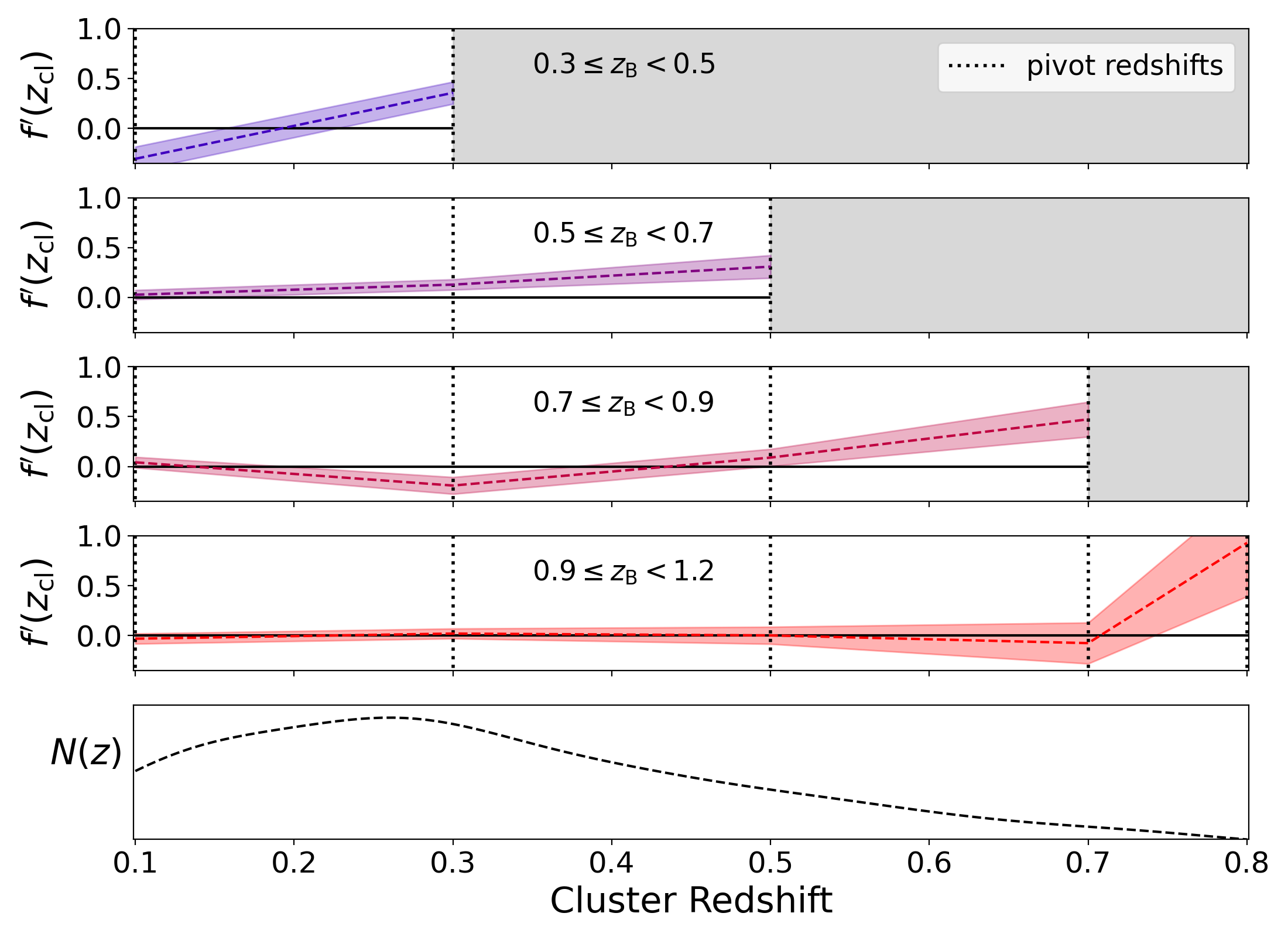}
    \caption{Best-fit result of the amplitude parameter of the global contamination model at pivot points in the cluster redshift space. Bottom: Distribution of the eRASS1 KiDS-1000 cluster redshifts.}
    \label{fig:f_zcl}
\end{figure}

Fig. \ref{fig:f_zcl} shows the amplitude parameter of the contamination model $f^\prime(z_\mathrm{cl})$ at pivot points  in the cluster redshift space. We observe that the highest redshift tomographic bin ($0.9<z_\mathrm{B}<1.2$) is effectively contamination-free for cluster redshifts below 0.7. We expect this behaviour, as the contamination should increase for clusters that approach the lower border of the tomographic bin in redshift space. There are no clusters with redshift above 0.8 in the eRASS1 cluster sample, and the highest redshift tomographic bin only starts at a photo-$z$ of 0.9.

The tomographic bin $0.7<z_\mathrm{B}<0.9$ shows significant contamination for $z_\mathrm{cl}>0.5$, which we expect due to analogous reasoning. However, the majority of eRASS1 clusters is located at lower cluster redshifts.

The tomographic bin $0.5<z_\mathrm{B}<0.7$ starts to show more contamination at lower cluster redshifts. We trace this back to the individual redshift distributions of the tomographic bins (Fig. \ref{fig:tbins}). This particular tomographic bin shows a pronounced wing towards lower redshifts. I.e. a large portion of galaxies in this photo-$z$ range populate lower actual redshifts, which leads to a stronger contamination.

The low-redshift tomographic bin ($0.3<z_\mathrm{B}<0.5$) only shows contamination for $z_\mathrm{cl}>0.2$, and negative contamination for cluster redshifts that are lower. This is a hint that our image injection analysis did not capture the full effect that cluster galaxies have on background galaxies in this tomographic bin, at least for very low cluster redshifts.

As a general rule we observe that cluster member contamination starts to become pronounced if the redshift separation between the cluster and tomographic bin is $\Delta z\lesssim 0.2$.\\

\begin{figure*}
    \centering
    \includegraphics[width=\textwidth]{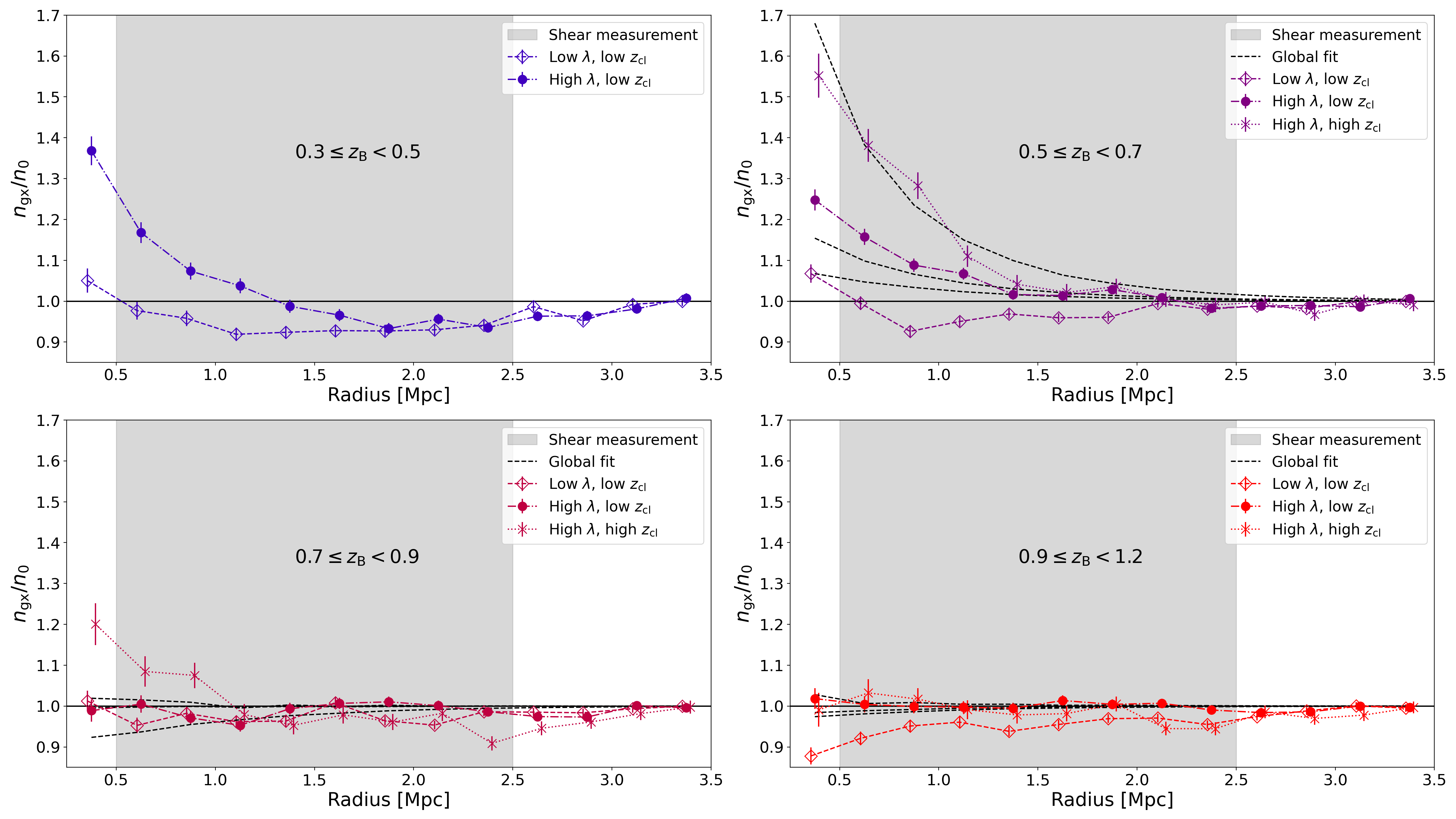}
    \caption{Measured and best-fit detection-corrected radial number density profiles, stacked in tomographic bins (different panels) and divided into three cluster sub-samples. We employ inverse-variance weighting to take the average at each radius, which pronounces clusters scattering low.}
    \label{fig:cont_model}
\end{figure*}

We proceed similarly to the stacking method that we described previously for the global fit of the detection profiles to check our global fit of the boosted number density profiles. We re-use the three cluster sub-samples divided by cluster redshift \& richness, but employ inverse variance weighting for the individual normalised, corrected number densities in each annulus. We 
%transfer
propagate
these inverse variance weights when we calculate the average of the corresponding model values. We chose a weighted average this time, because the unweighted average will be biased high due to some values that up-scatter significantly (but have large absolute uncertainties and therefore a low impact on the global fit).

We note that this method of stacking (using the inverse-variance weights) will emphasize number density profiles that scatter low. This is due to the Poisson errors; data points with few galaxies will have a low absolute uncertainty and subsequently a high weight contribution to the average. This weighted average does, however, allow us to verify if the global fit describes the data. We show the result in Fig. \ref{fig:cont_model}.

As we discussed previously it becomes evident that the image injection analysis, which we use to account for the object detection bias, does not fully capture the effect in the lowest tomographic bin ($0.3<z_\mathrm{B}<0.5$). We observe a dip in the stacked profiles below the baseline, before the contamination dominates at small cluster-centric distances. While some of this can be attributed to the inverse-variance weights of the stacked number density profiles, we do not observe such a behaviour to this extent in the other tomographic bins.

This, together with the redshift evolution of the amplitude parameter of the contamination model in this tomographic bin (Fig. \ref{fig:f_zcl}), is enough reason for us to not use the shear measurement in this tomographic bin in our further analysis and mass calibration. This tomographic bin has a low effective lensing efficiency, hence we do not loose a lot of lensing signal by disregarding it. On the other hand we want to 
%secure
ensure
that the cluster member contamination correction is as accurate as possible, as we will perform consistency checks with DES \& HSC and want to provide accurate shear measurements for the eRASS1 WL mass calibration.

As expected from its redshift distribution (Fig. \ref{fig:tbins}), the tomographic bin $0.5<z_\mathrm{B}<0.7$ shows the most contamination. The stacked number density profiles in the two high-redshift tomographic bins are 
%essentially
nearly contamination free (the tomographic bin $0.7<z_\mathrm{B}<0.9$ shows some contamination in the high-redshift, high-richness clusters sub-sample at radii below $\SI{1}{\mega pc}$).

\begin{figure}
    \centering
    \includegraphics[width=\linewidth]{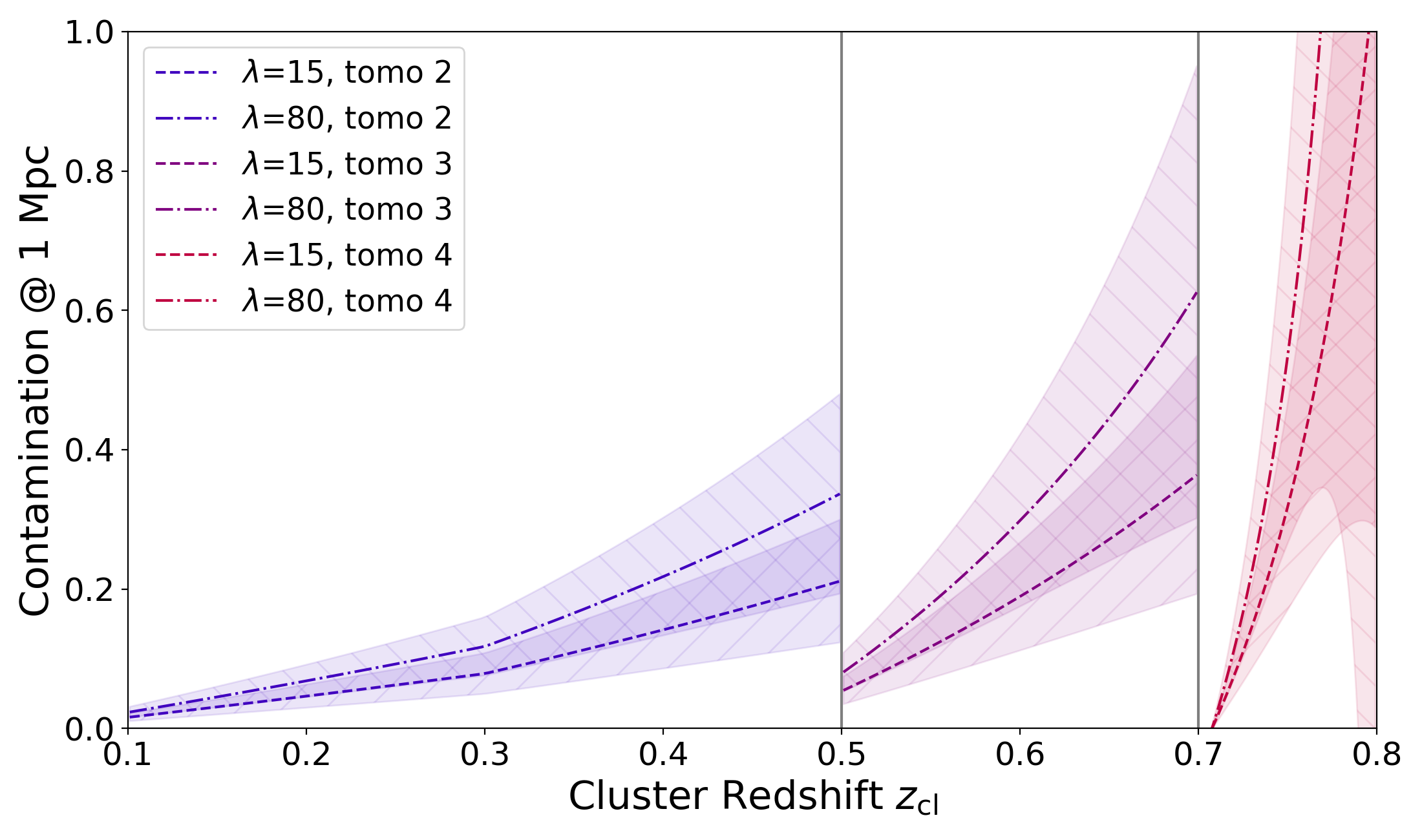}
    \caption{Best-fit contamination model for richness 15 and 80 at a cluster-centric distance of $\SI{1}{\mega pc}$ as a function of the cluster redshift, and always showing the lowest included tomographic source bin only.}
    \label{fig:cont_demo}
\end{figure}

We show the best-fit contamination model at a cluster-centric distance of $\SI{1}{\mega pc}$ for two cluster richnesses ($\lambda=80$ and $\lambda=15$), as a function of the cluster redshift in Fig. \ref{fig:cont_demo}.\\
To account for the contamination in the further analysis, one can either rescale the individual shear profiles or include the contamination in the computation of the weak lensing mass bias $b_\mathrm{WL}$.  We follow the latter approach for our main analysis, which allows us to easily propagate the uncertainty in the fit of the global contamination model (see Appendix \ref{app:bwl}).

%We decide against using the global contamination model to boost the shear profiles directly, as the propagation of error would have decreased the signal-to-noise of the noisy individual shear profiles even further. Instead, we include the contamination model in the calculation of the mass bias parameter $b_\mathrm{WL}$ (App. \ref{app:bwl}).

\section{Consistency checks with DES Y3 and HSC S19}
\label{sec:consistency}

In this section we compare the WL results obtained on eRASS1-selected clusters using the KiDS-1000 survey, the Dark Energy survey \citep{Grandis_2023}, and the HSC surveys \citep[Section 2.3.1]{ghirardini23}. We will perform this comparison both at the level of individual density contrast estimates \tim{and  amplitudes}, as well as at the level of the global scaling relation fit.

\begin{figure*}
    \centering
   \includegraphics[width=\columnwidth]{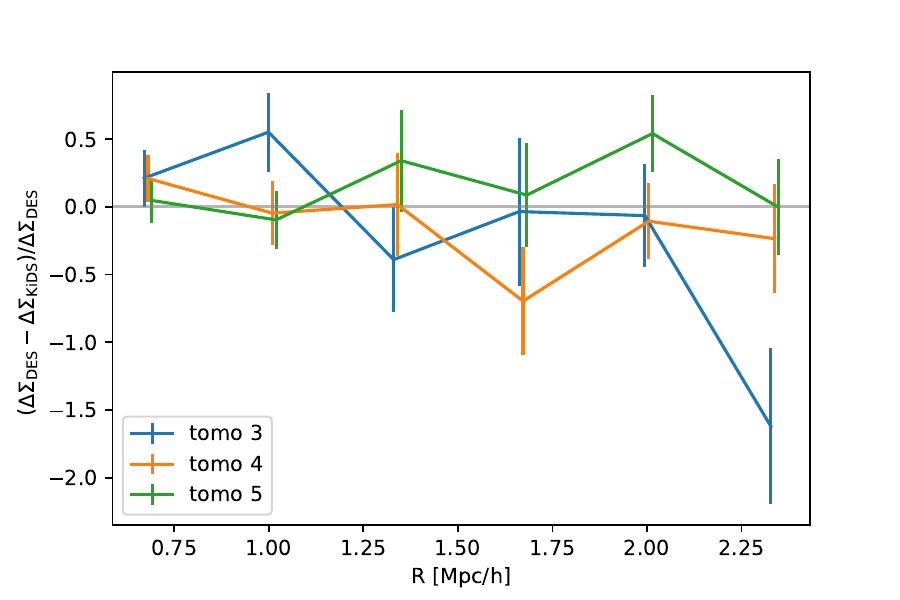}
   \includegraphics[width=\columnwidth]{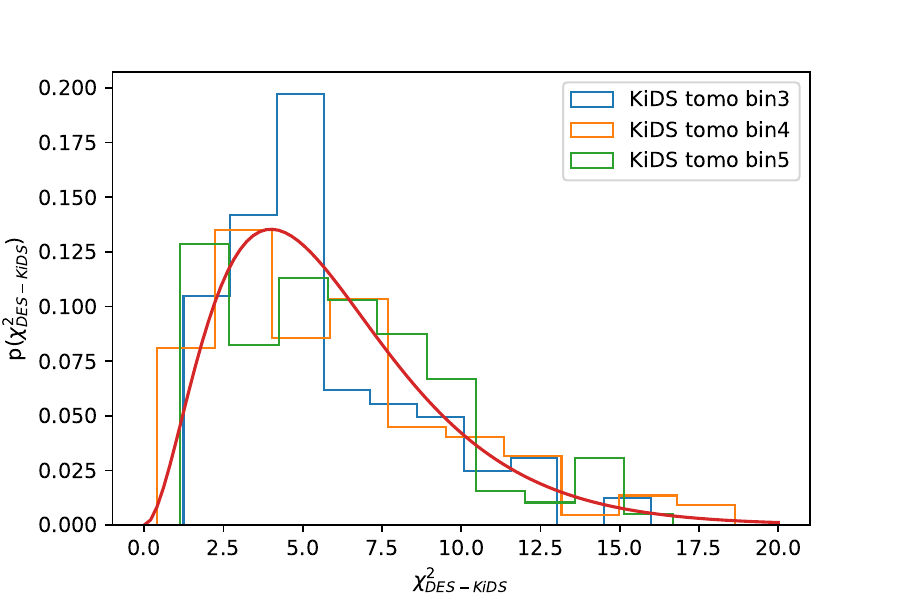}
    \includegraphics[width=\columnwidth]{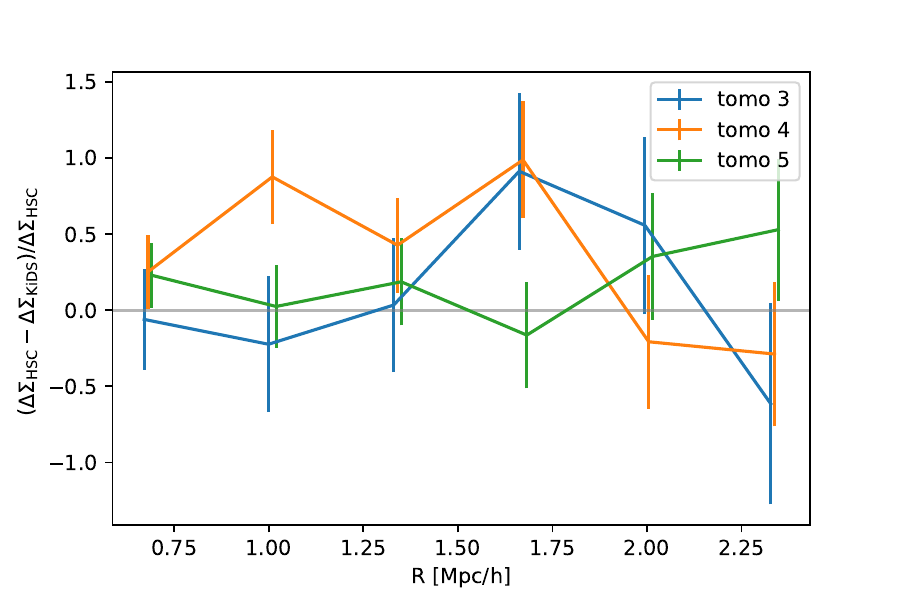}
   \includegraphics[width=\columnwidth]{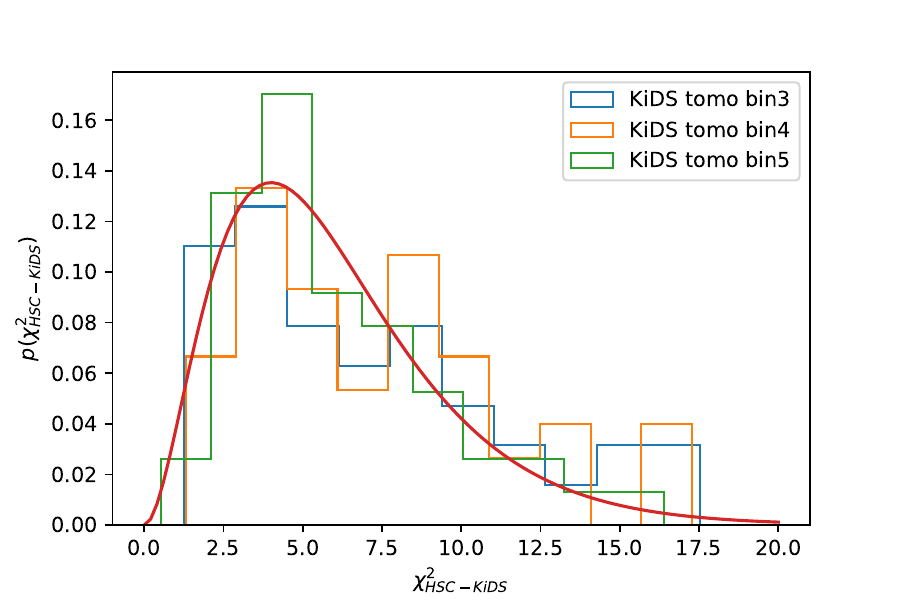}
  \caption{Consistency test between the density contrast profiles of clusters observed both in DES (HSC) and KiDS in the upper (lower) panels. Given the correlated intrinsic galaxy shapes, we construct the statistical uncertainty estimates by bootstrapping on the union of the respective source catalogs. We find that, averaged over all objects, the estimates from the survey pairs are consistent as a function of radius (left column). Also, the uncertainty weighted mean square deviation of all clusters, shown in the right column, is consistent with the expectation (red curve).} \label{fig:indiv_consistency}
\end{figure*}

\subsection{Contamination-corrected shear profiles}\label{sec:comp_DeltaSigma}

Within the western galactic hemisphere, the KiDS-1000 survey has a footprint overlap both with the DES survey, as well as with the HSC survey (Fig. \ref{fig:footprint}). The direct overlap between HSC and DES falls in the eastern galactic hemisphere, and thus outside of the region for which the eROSITA-DE consortium is analysing the X-ray data. The joint overlap with both surveys provides the unique opportunity to validate cluster WL measurements from different lensing surveys around the lens clusters selected in a coherent manner.

Given the different source redshift distributions and the different degrees of cluster member contamination, this comparison 
is best conducted on the 
%can only reasonably take place on the %estimated density contrast profiles 
estimates for the  density contrast profiles 
$\Delta\Sigma(R)=\Sigma_{\text{crit,eff}} \, \langle\gamma_\mathrm{t}\rangle(R)$.
Here, $\langle\gamma_\mathrm{t}\rangle(R)$ is estimated from the contamination-corrected azimuthally averaged tangential shear estimates\footnote{He we approximate the shear with the reduced shear and ignore corrections from the convergence (see Eq.\thinspace\ref{eq:reducedshear}) given the lack of a precise convergence profile estimate for each cluster. However, reduced shear corrections are correctly accounted for in the more constraining scaling relation comparison \citep[see Sect.\thinspace\ref{se:scaling_relations} and][]{Grandis_2023}.}, while 
differences in the source redshift distributions are accounted for via the effective critical surface mass density $\Sigma_{\text{crit,eff}}$, which is computed 
by adequately averaging  (see below)  
the source and lens redshift-dependent inverse 
%as an adequately weighted source-population average (see below) of the 
critical 
 surface mass density
\begin{equation}
\label{eq:sigma_crit}
    \Sigma^{-1}_\mathrm{crit,ls}=\frac{4\pi G}{c^2}  \frac{D_\mathrm{l}D_\mathrm{ls}}{D_\mathrm{s}} H(z_\mathrm{s}-z_\mathrm{l})  \,.
\end{equation}
%with 
Eq.\thinspace\ref{eq:sigma_crit} includes the
 the Heaviside step function $H(x)$,  the gravitational constant $G$, the speed of light in vacuum $c$, and the angular diameter distances between the observer and the lens $D_\mathrm{l}$,  the observer and the source $D_\mathrm{s}$, and the lens and the source $D_\mathrm{ls}$.
%over the source population (see below) to account for the differences in the redshift distributions of the different surveys.

%which is defined via the true azimuthally averaged tangential shear  $\gamma_\mathrm{t}$ (thus requiring a contamination correction in the estimation) and the average critical surface mass density $\Sigma_{\text{crit}}=...$

Comparing the density contrast estimates of the same lenses measured in two WL surveys poses the challenge that both surveys will have selected some source galaxies in common. The shape noise contributions from these sources will be partially correlated, as the intrinsic shape of a common source is the same\footnote{Slight differences may occur due to different resolution in different data sets and different brightness profile weighting schemes employed by different shape measurement techniques.}, while the shape measurement error will be different. Working only on sources selected by both surveys is in principle possible \citep{Hoekstra_2015}, but would require to re-calibrate shape measurement and photo-z uncertainties for the new, joint selection, as source redshift and shape calibrations are only valid for specific survey-internal source selections. These selection criteria would be broken when working on a jointly selected background galaxy sample.

To circumvent this problem we construct the \emph{union} of the source catalogs from KiDS and DES around each eRASS1 clusters (and KiDS and HSC, respectively). Sources from the two surveys that fall within 1 arcsec of each other are considered selected by both surveys, and successively treated as one source with two sets of measurements. Naturally, other sources will be selected by only one of the surveys. We thus construct two union samples, DES-KiDS around 125 eRASS1 clusters, and HSC-KiDS around 48 eRASS1 clusters. Compared to previous cluster WL comparisons, our approach has the benefit of being able to rely on the high accuracy of shape and photo-$z$ calibration of the respective wide-area WL surveys.

We estimate the density contrast from the DES sources as
\begin{equation}
    \Delta\Sigma_\text{DES}(R) = \frac{1}{1-f_\text{cl}^\text{DES} (R | z_\text{l},\lambda)} \frac{ \sum_i w^{b_i} \Sigma_{\text{crit,l}b_i}  w^\mathrm{s}_i e_{\mathrm t,\,i}  }{\sum_i w^{b_i} w^\mathrm{s}_i \mathcal{R}_{i} },
\end{equation}
where $f_\text{cl}(R | z_\text{l},\lambda)$ is the best-fit cluster member contamination profile derived in \citet[Section 3.3.1]{Grandis_2023}, which is a function of cluster centric distance $R$, cluster richness $\lambda$ and cluster redshift $z_\text{l}$. The sum over $i$ runs over all selected DES sources $i$ in the respective radial bin, $w^{b_i}$ is the weighting of the sources' tomographic redshift bin $b_i$, which implements the DES background selection \citep[][Section 3.2]{Grandis_2023}. The critical density of the tomographic bin is computed as $\Sigma_{\text{crit,l}b}^{-1} = \int \text{d} z_\text{s} P(z_s | b) \Sigma_\text{crit,ls}^{-1}$, where $P(z_s | b)$ are the source redshift distributions calibrated by \citet{myles21}. Note that they are normalized to the multiplicative shear bias correction of that tomographic bin. Our implementation thus corrects for the multiplicative shear bias.

To estimate the surface mass density from the HSC source data, we use the source sample whose selection follows \citet{chiu22}, as reported in \citet[Section 2.3.1]{ghirardini23}. Specifically
\begin{equation}
     \Delta\Sigma_\text{HSC}(R) = \frac{1}{2 (1 + \mathcal{K}) \mathcal{R}} \frac{ \sum_i \Sigma_{\text{crit,l}i}  w^\mathrm{s}_i e_{\mathrm t,\,i}  }{\sum_i w^\mathrm{s}_i },
\end{equation}
where $\mathcal{K}=\sum_i w^\mathrm{s}_i m_i / \sum_i w^\mathrm{s}_i$ is the source-weighted average multiplicative shear bias and $\mathcal{R} = 1- \sum_i w^\mathrm{s}_i e_{\text{rms},i}^2 / \sum_i w^\mathrm{s}_i $. In light of the stringent background selection implemented, the HSC WL measurements have a cluster member contamination that is consistent with zero. 
%For the critical  density of each source, $\Sigma_{\text{crit,l}i}$, we employ a Monte Carlo realisation of the \texttt{DEmP} photo-$z$ estimates \citep{Hsieh_2014}.
For the HSC analysis we compute individual estimates for the critical surface mass density  $\Sigma_{\text{crit,l}i}$ for each source $i$ from a Monte Carlo realisation of \texttt{DEmP} photo-$z$ estimates \citep{Hsieh_2014}, which yield unbiased estimates of  $\Sigma_{\text{crit}}$ at the $\lesssim 1\%$ level out to cluster redshifts $z_\mathrm{cl}\sim 0.6$ \citep{chiu22}.

For the KiDS data, we estimate the density contrast for each tomographic bin $b$ as
\begin{multline}
    \Delta\Sigma_{\text{KiDS},b}(R) = \\ = \frac{1}{1-f_\text{cl}^\text{KiDS} (R | z_\text{l},\lambda,b)} \frac{1}{\langle \Sigma_{\text{crit},b}^{-1} \rangle} \frac{1}{1+ m_b}   \frac{ 
%    \sum_i  w^\mathrm{s}_i e_{\mathrm t,\,i}  }{\sum_i w^\mathrm{s}_i },
   \sum_i  w^\mathrm{s}_i \epsilon_{\mathrm t,\,i}  }{\sum_i w^\mathrm{s}_i },
\end{multline}
with $f_\text{cl}^\text{KiDS} (R | z_\text{l},\lambda,b)$ being the cluster member contamination model of each tomographic bin,  $\langle \Sigma_{\text{crit},b}^{-1} \rangle$ constituting the average inverse critical surface mass density of that tomographic bin, and $m_b$ being the mean multiplicative shear bias of the respective bin, listed in Table~\ref{tab:KiDS-multbias}.

On an individual cluster basis, our test statistic is the difference between the DES and KiDS density contrasts
\begin{equation}
    \Xi_{\text{DES-KiDS},b}(R) = \Delta\Sigma_\text{DES}(R) - \Delta\Sigma_{\text{KiDS},b}(R),
\end{equation}
for each KiDS tomographic bin $b$, and the difference between the HSC and KiDS density contrasts
\begin{equation}
    \Xi_{\text{HSC-KiDS},b}(R) = \Delta\Sigma_\text{HSC}(R) - \Delta\Sigma_{\text{KiDS},b}(R).
\end{equation}
Statistical errors $\delta \Xi(R)$ on these profiles need to be extracted by bootstrapping on the unions of the respective source samples. Specifically, this means that if in a bootstrap we draw a source present in \emph{both} surveys, it needs to be coherently included in both parts of the difference. Note also, that by bootstrapping on the union of the samples, we are still bootstrapping on a fair sample of either surveys selection. This means that selection-dependent calibration products are still valid.

In Fig.~\ref{fig:indiv_consistency} we show the sample-averaged difference as a function of cluster-centric distance in the reference cosmology in the left panels, and the inverse-variance-scaled squared deviation of our test statistics from zero
\begin{equation}
    \chi^2 = \sum_R \frac{ \Xi(R)^2}{\delta \Xi(R)^2 },
\end{equation}
in the right panels.
The latter are compared with the expected distributions if $\Xi(R) \sim \mathcal{N}(0, \delta \Xi(R)^2)$, that is from a random normal with mean zero, and variance given by the square of the bootstrap error we determine. This is a chi-squared distribution with 6 degrees of freedom, as we choose 6 radial bins. Compressing the left panels of  Fig.~\ref{fig:indiv_consistency} further into a single global number by averaging over KiDS tomographic bins and cluster-centric distances, we find relative differences of $-0.117 \pm 0.066$ (1.8 $\sigma$) and $0.212 \pm 0.082$ (2.6 $\sigma$) between DES versus KiDS and HSC versus KiDS, respectively, both statistically consistent with zero. The significances in the individual tomographic bins of KiDS are smaller due to the increased noise.

\begin{figure*}
    \centering
   \includegraphics[width=0.49\textwidth]{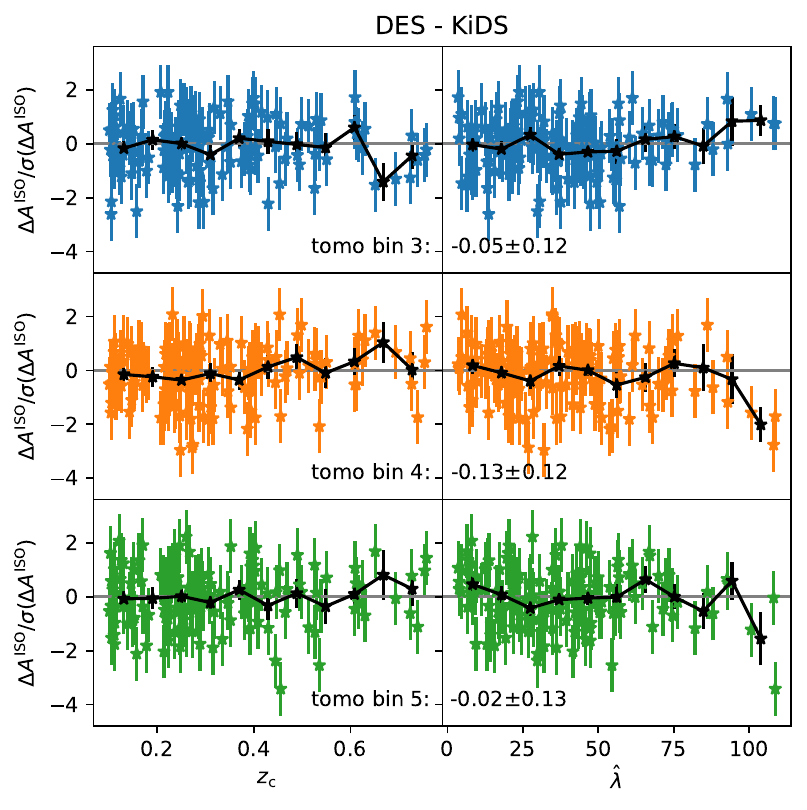}
   \includegraphics[width=0.49\textwidth]{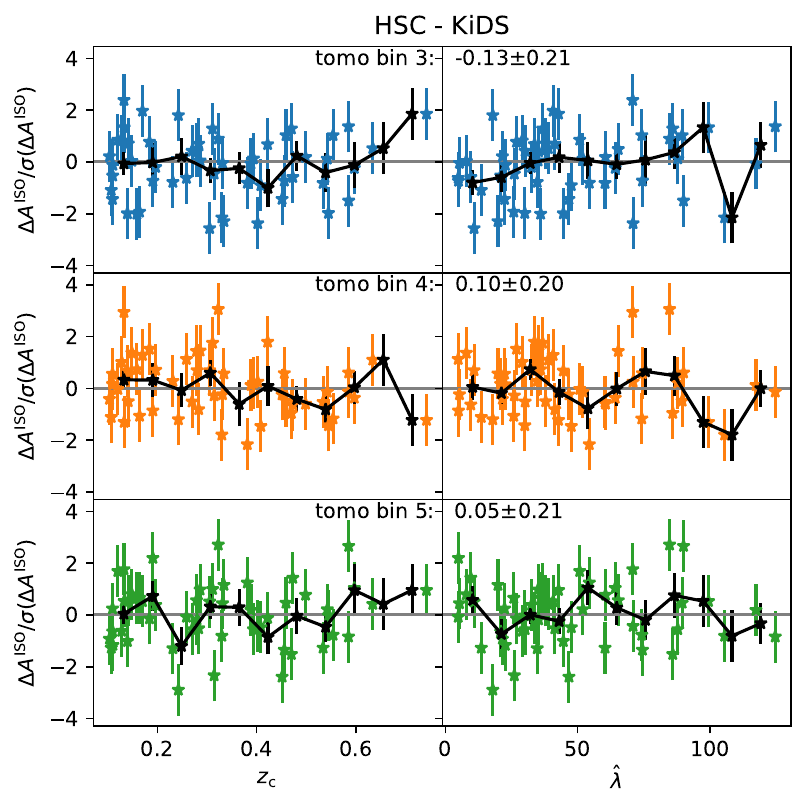}
  \caption{Difference in the amplitude of isothermal sphere \tim{shear profile models} fitted to clusters with coverage by DES and KiDS (HSC and KiDS) in the left (right) \tim{panel}, plotted \tim{relatively to the uncertainty of this difference and} as a function of cluster redshift and richness (columns) for the different KiDS tomographic redshift bins (rows). Excellent overall agreement is found, while no richness or redshift trends are evident from visual inspection, \tim{also} after binning (black points).  }  \label{fig:SR_consistency}
\end{figure*}

\subsection{\tim{Scaled }Einstein radius comparison}
To test for redshift and richness trends in the WL signal of clusters observed by two surveys, we expand on the Einstein radius comparison proposed in \citet{Hoekstra_2015}. That comparison was based on the fact that for a mass distribution following an isothermal sphere, the tangential shear profile takes the form $\gamma_\text{t}(R) = \frac{R_\text{E}}{2 R}$, where $R_\text{E}$ is the mean Einstein radius of the source sample. Ignoring the reduced shear correction, it can be directly \tim{obtained using}
%estimated from the with 
an estimator of the form $\langle 2 \epsilon_t R \rangle$, thus providing a convenient measure of the overall strength of the WL signal around a lens, avoiding to split in radial bins, as done in Section~\ref{sec:comp_DeltaSigma}.

In order to account for the different source redshift distributions of our source samples, we alter the original prescription by \citet{Hoekstra_2015} to estimate the amplitude $A^\text{ISO}$ of an isothermal density profile in physical units, which we define as
\begin{equation}\label{eq:iso_model}
    \gamma_\text{t}(\theta) = \Sigma_\text{crit}^{-1} A^\text{ISO} \frac{1}{\theta},
\end{equation}
where $\theta$ is the cluster centric distance in angular units. Direct comparison to the expression from \citet{Hoekstra_2015} reveals that $A^\text{ISO}$ is just a critical surface density-corrected Einstein radius in angular units, which can therefore be readily compared between source samples with different geometrical lensing configurations.

To estimate this quantity in the three different surveys we consider, we set up the following inverse variance weigthed estimators. For DES, we compute

\begin{equation}
    A^\text{ISO}_\text{DES} =  \frac{ \sum_i w^{b_i} w^\mathrm{s}_i \Sigma_{\text{crit,l}b_i}^{-1} (1-f_{\text{cl},i}^\text{DES}) \theta_i^{-1}  e_{\mathrm t,\,i}  }{\sum_i w^{b_i} w^\mathrm{s}_i \Sigma_{\text{crit,l}b_i}^{-2} (1-f_{\text{cl},i}^\text{DES})^2 \theta_i^{-2}\mathcal{R}_{i} },
\end{equation}
where $f_{\text{cl},i}^\text{DES} = f_\text{cl}^\text{DES} (R_i | z_\text{l},\lambda)$ is the cluster member contamination evaluated for the $i$th source. Note that this estimator \tim{scales the tangential reduced shear of each source $i$} 
%effectively weighs each sources $i$ tangential shear 
with $\Sigma_{\text{crit,l}b_i}  \theta_i/(1-f_{\text{cl},i}^\text{DES})$, in line with the naive inversion of Eq.\thinspace\ref{eq:iso_model} in presence of cluster member contamination.

For HSC, where \citet{chiu22} found no cluster member contamination, and where we have reliable individual photometric redshift estimates, we compute

\begin{equation}
    A^\text{ISO}_\text{HSC} =  \frac{1}{2 (1 + \mathcal{K}) \mathcal{R}} \frac{ \sum_i w^\mathrm{s}_i \Sigma_{\text{crit,l}i}^{-1} \theta_i^{-1}    e_{\mathrm t,\,i}  }{\sum_i w^\mathrm{s}_i \Sigma_{\text{crit,l}i}^{-2} \theta_i^{-2} }.
\end{equation}

Similarly, for each KiDS tomographic redshift bin $b$, we can compute

\begin{equation}
    A^\text{ISO}_{\text{KiDS},b}= \frac{1}{\langle \Sigma_{\text{crit},b}^{-1} \rangle} \frac{1}{1+ m_b}   \frac{ 
   \sum_i  (1-f_{\text{cl},bi}^\text{KiDS}) \theta_i^{-1} w^\mathrm{s}_i \epsilon_{\mathrm t,\,i}  }{\sum_i w^\mathrm{s}_i  (1-f_{\text{cl},bi}^\text{KiDS})^2 \theta_i^{-2} },
\end{equation}
with the cluster member contamination estimate $f_{\text{cl},bi}^\text{KiDS}= f_\text{cl}^\text{KiDS} (R_i | z_\text{l},\lambda,b)$ for each individual source. 

As test statistic, we then consider the difference between the amplitude estimates in HSC and KiDS, as well as in DES and KiDS for the clusters that have data in two of the respective surveys. The statistical uncertainty on these differences is estimated by bootstrapping on the \emph{union} of the source samples, as discussed in Section~\ref{sec:comp_DeltaSigma}.

\begin{table}
    \centering
    \caption{Sample averaged differences in the amplitudes of \tim{scaled} isothermal sphere \tim{shear profiles} relative to the uncertainty of those differences measured in DES and HSC w.r.t. to KiDS estimates based on different tomographic redshift bins.}\label{tab:iso-check}
    \begin{tabular}{lccc}
         tomo bin & $b=3$ & $b=4$ & $b=5$ \\
         \hline
         DES & $-0.05 \pm 0.12$ & $ -0.13 \pm 0.12$ & $-0.02 \pm 0.13$ \\
         HSC & $-0.13 \pm 0.21$ & $0.10 \pm 0.20$ & $0.05 \pm 0.21$ \\
    \end{tabular}
\end{table}

\begin{figure*}
    \centering
   \includegraphics[width=\textwidth]{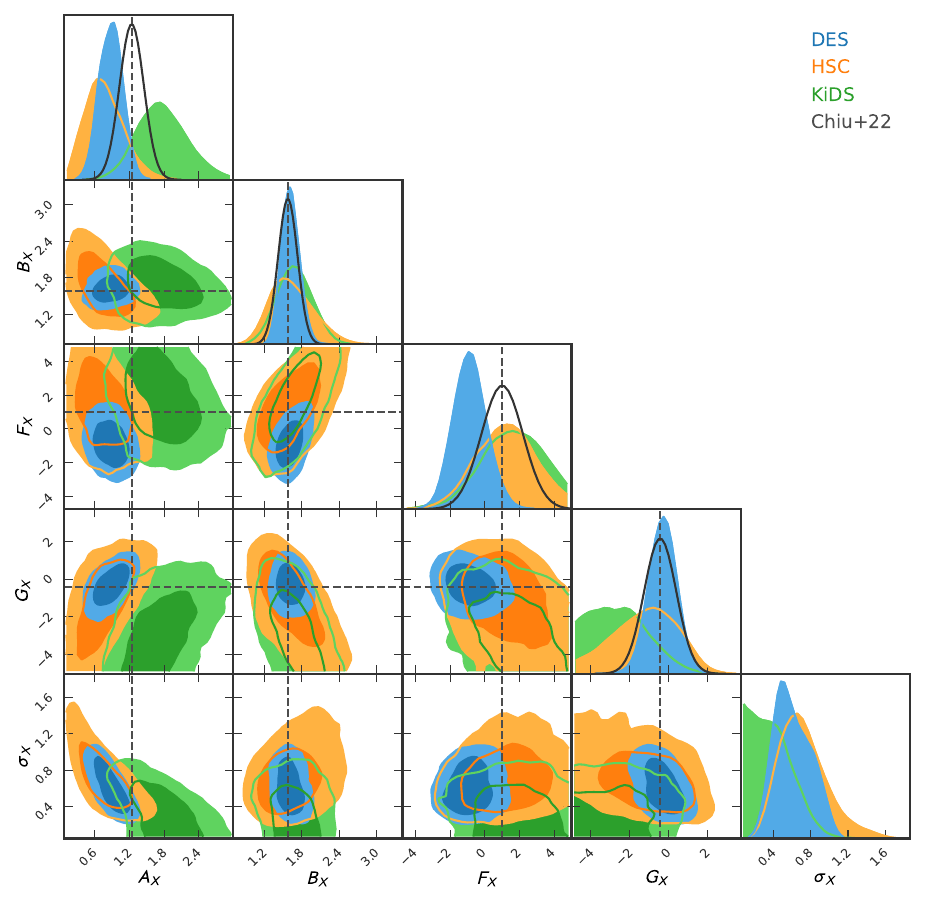}
  \caption{1- and 2-dimension marginal contour plot for the WL mass calibration of the eROSITA count rate to halo mass relation. In green the results from KiDS-1000 on eRASS1 are shown, in blue the results from DES Y3 on eRASS1 \citep{Grandis_2023}, in orange the results from HSC-Y3 on eRASS1, and in black the results from HSC-Y3 on eFEDS \citep{chiu22}. All results show agreement, indicating the reliability of the WL mass calibration methods employed by the different teams. }  \label{fig:iso_comp}
\end{figure*}

In Fig.~\ref{fig:iso_comp} we present in the left (right) \tim{panel} the 
%distributions with cluster redshift and richness
\tim{distribution}
of the difference in isothermal sphere amplitudes between DES and KiDS (HSC and KiDS) for the different KiDS tomographic redshift bins,
\tim{plotted relatively to the uncertainty of this difference as a function of cluster redshift and richness}. In black  \tim{the figure also shows averages computed over equally spaced bins including the respective error bars}. We find excellent agreement between these estimates, with the overall mean also reported in Table~\ref{tab:iso-check}. Furthermore, visual inspection reveals no redshift and richness trends. 

Compared to the test in the previous section, this test places more focus on the outer parts of the profile due to the $\theta$ weighting in the estimator. This regime is less impacted by cluster member contamination and obscuration of background sources, while focussing on the scales where the bulk of the signal originates. It is therefore very encouraging to find such a good agreement among the estimates from the different surveys.

\subsection{Scaling relations}
\label{se:scaling_relations}

Using the Bayesian population model described in \citet[Section 5.1]{Grandis_2023} and \citet[Section 4]{ghirardini23} we derive constraints on the parameters of the scaling between the eROSITA count rate, and halo mass and redshift, defined as follows
\begin{equation}
\bigg\langle\log \frac{C_R}{C_{R,\text{b}}} \bigg| M, z\bigg\rangle = 
\log A_\text{X} + 
b_X(M, z) \cdot \log \frac{M}{M_\text{b}} + e_x(z)\,,
\label{eq:CR_given_M_z}
\end{equation}
where $C_{R,\text{b}} = 0.1$ is the pivot value for the count rate, $M_\text{b} = \SI{2e14}{\mathrm{M}_\odot}$ is the pivot value for the mass, and $z_\text{b} = 0.35$ is the pivot value for the redshift. $b_X(M, z)$ expresses the mass-redshift dependent slope of the scaling relation, given by
\begin{equation*}
b_X(M, z) = \bigg( B_X + C_X \cdot \log \frac{M}{M_\text{b}} + F_X \cdot \log \frac{1+z}{1+z_\text{b}} \bigg)
\end{equation*}
where $B_X$ is the classic standard single value mass slope, $C_X$ allows the slope to be mass dependent -- but is kep fixed to zero in this analysis, and $F_X$ allows the slope to be redshift dependent. 

The redshift evolution term, $e_x(z)$, is given by

\begin{equation}
e_x(z) = D_X \, \log \frac{d_L(z)}{d_L(z_\text{b})} + E_X \, \log \frac{E(z)}{E(z_\text{b})} + G_X \, \log \frac{1+z}{1+z_\text{b}}
\label{eq:e_xz}
\end{equation}

\noindent for which $E_X = 2$ and $D_X=-2$ are the values adopted by the self-similar model \citep{Kaiser1986} for the case of luminosity.

As described in more detail in \citet{Grandis_2023}, Section 5.1, and \citet{ghirardini23}, Section 4, we construct a likelihood for the measured  tangential reduced shear profile. Deviating from the other works, for the KiDS data, we consider \tim{one} reduced shear profile $\hat g_{\text{t},b}$ for \tim{each} tomographic redshift bin $b$. The likelihood depends not only on the parameters of the count rate to halo mass relation, but also on the cosmological parameters. We thus vary the present day matter density w.r.t. to the critical density, $\Omega_\text{M} = 0.331 \pm 0.038$, as determined by the Dark Energy Surveys year 1 supernovae type Ia analysis \citep{Abbott_2019}. We also vary the fraction of random sources and point sources in our likelihood with priors from the optical follow-up of eRASS1 clusters \citep{kluge23, ghirardini23}.

Another input is the calibration of the WL mass bias and scatter, which follows the prescriptions outlined in \citet[Section 4]{Grandis_2023}. The adaptations to the KiDS-1000 WL specifications, and resulting numerical values for the WL bias and scatter are discussed in App.~\ref{app:bwl}.

\begin{table*}
\caption{\label{tab:SR}
Calibration of the count rate to halo mass relation parameters %of eRASS1 
\tim{for eROSITA} using different weak lensing data sets and cluster samples.}
\begin{tabular}{l|c|c|c|c|c}
WL survey & $A_\text{X}$ & $B_\text{X}$ & $F_\text{X}$ & $G_\text{X}$ & $\sigma_\text{X}$\\
\hline\hline
KiDS-1000 (eRASS1) & $1.77\pm 0.42$ & $1.68\pm 0.27$ & $1.57\pm 1.74$ & $-2.60 \pm 1.45$ & $<0.85$ \\
DES Y3 (eRASS1) & $0.88\pm 0.20$ & $1.62\pm 0.14$ & $-0.85\pm 0.93$ & $-0.32 \pm 0.69$ & $0.61\pm 0.19$ \\
HSC-Y3 (eRASS1) & $0.77\pm 0.32$ & $1.65\pm 0.34$ & $1.20\pm 1.66$ & $-1.24 \pm 1.60$ & $0.72\pm 0.25$ \\
HSC-Y3 (eFEDS) & $1.24\pm 0.21$ & $1.58\pm 0.16$ & $1.00\pm 1.20$ & $-0.44 \pm 0.83$ & -- \\
\multicolumn{6}{p{\linewidth-12pt} }{\small Constraints on the parameters of the count rate to halo mass relation for eROSITA selected clusters, derived from different WL surveys (KiDS, DES, HSC), and on different eROSITA surveys (eRASS1, eFEDS). All of these analyses are statistically consistent with each other. }
\end{tabular}
\end{table*}

Sampling the likelihood we find the amplitude of the count rate to halo mass relation \mbox{$A_\text{X}=1.77\pm 0.42$}, the mass trend of the 
%count rate mass
relation \mbox{$B_\text{X}=1.69\pm 0.27$}, its redshift trend \mbox{$F_\text{X}=1.57\pm 1.74$}, the deviation from a self-similar redshift evolution \mbox{$G_\text{X}=-2.60 \pm 1.47$}, and an upper limit on the intrinsic scatter of \mbox{$\sigma_\text{X}<0.85$} at the 97.5 percentile. The 1- and 2-dimensional marginal contours of the posterior are shown in green in Fig~\ref{fig:SR_consistency}, and listed in Table~\ref{tab:SR}.

We repeat the same analysis also with the HSC derived WL data set, extracted with the method described in \citet{chiu22}, and outlined in \citet[Section 2.3.1]{ghirardini23}. The WL bias and scatter for HSC are reported in Appendix \ref{app:bWL_hsc}. With that data set we find an amplitude of the observable--mass relation \mbox{$A_\text{X}=0.77 \pm 0.32$}, a mass slope \mbox{$B_\text{X}=1.65\pm 0.35$}, its redshift trend \mbox{$F_\text{X}=1.20\pm 1.66$}, a deviation from self-similar redshift evolution of \mbox{$G_\text{X}=-1.24\pm 1.60$}, and an intrinsic scatter of \mbox{$\sigma_\text{X}=0.72\pm 0.25$}. These  results are shown as orange contours in Fig.~\ref{fig:SR_consistency} and Table~\ref{tab:SR}.

Given that all three mass calibration analyses have coherent WL bias correction, and calibrate the same astrophysical relation between eROSITA count rate and halo mass, we can directly compare the results on the scaling relation parameters from KiDS and HSC with the DES results \citep[see Fig.\thinspace\ref{fig:SR_consistency} in blue and][]{Grandis_2023}. 
%but also in Fig. \ref{fig:SR_consistency} in blue), 
We find that the mass calibration from all three surveys is statistically consistent with each other, and the results found by \citet{chiu22} with HSC on eFEDS clusters. We quantify this agreement by following the prescription of \citet{raveri+21} based on the posterior of parameter differences, and the probability mass of a difference consistent with zero. We find that in our case, the Gaussian limit holds for the posterior of parameter differences. Turning this probability mass into a tension in sigmas, we find $1.33\sigma$ of tension between the scaling relation parameter posteriors from HSC and KiDS (without eFEDS), $2.7\sigma$ between KiDS and DES, and $0.23 \sigma$ between DES and HSC. The analyses in the three surveys have been performed on different data sets, with different modeling choices. Their agreement validates these choices. It demonstrates reliability of WL mass calibration of clusters in the context of Stage III surveys.

\subsection{Goodness of fit}

\begin{figure*}
    \centering
   \includegraphics[width=\textwidth]{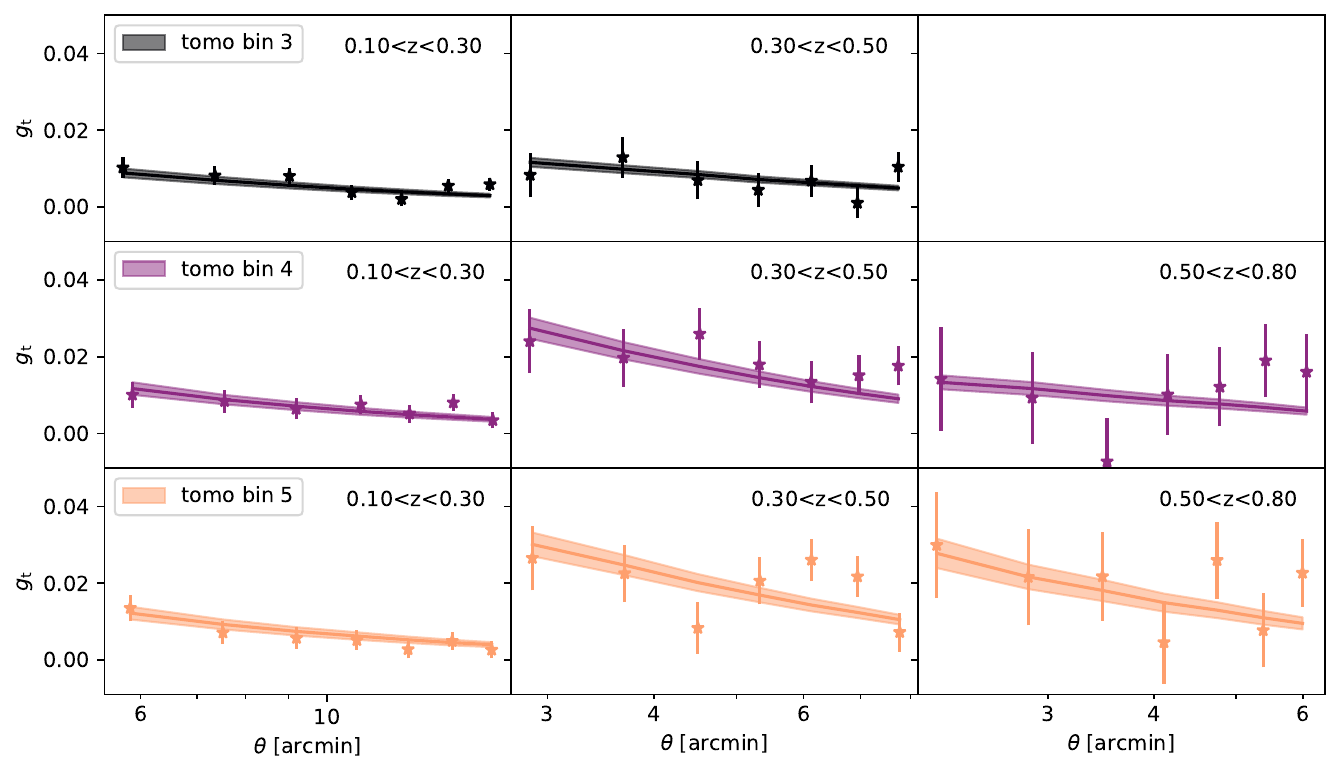}
  \caption{Stacked tangential reduced shear profiles from KiDS-1000 for different tomographic redshift bins (rows), and cluster redshift bins (columns), together with the mean stacked model (which incorporates the model for cluster member contamination, making the model prediction and the data directly comparable) and its one sigma uncertainty, as filled area. The model fits the data very well.}  \label{fig:kids_goodnessfofit}
\end{figure*}

\begin{figure*}
    \centering
   \includegraphics[width=\textwidth]{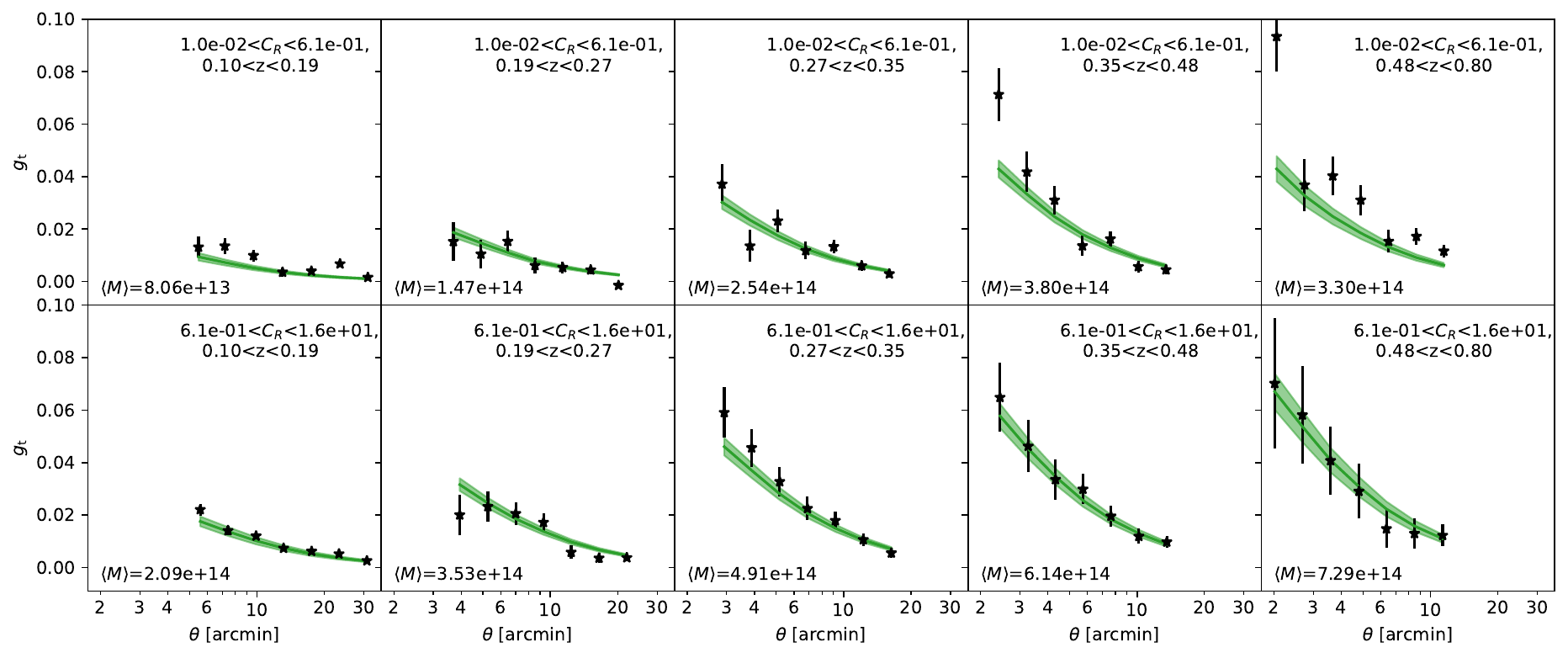}
  \caption{Stacked tangential reduced shear profiles measured using the HSC data and split according to cluster count rate (rows), and cluster redshift bins (columns), together with the mean stacked model and its one sigma uncertainty, as filled area. Not all stacked reduced shear profiles result in statistically good fits (especially at low count rates), leading to rather high $\chi^{2}$.}  \label{fig:HSC_goodnessoffit}
\end{figure*}

To inspect the goodness of fit, we follow the same procedure as in \citet[Section 5.2]{Grandis_2023}. We refer the reader to that section for the exact definitions of the stacking procedure. Differing from that work, for the KiDS data, we define the stacking weight for the cluster $\varkappa$  as $W_{\varkappa, b} = \sum_i w^{s}_{i,\varkappa,b} $ for each tomographic bin $b$ in KiDS. We also only consider three redshift bins (0.1, 0.3, 0.5, 0.8), and no binning in count rate. The rest of the procedure is the same. Fig. 15 shows that the jointly constrained model describes the measured tangential reduced shear profiles well for all KiDS tomographic redshift bins, with a chi-squared of $\chi^{2}_\text{KiDS} = 46.0^{+9.1}_{-0.4}$ with 63 data points, 3 well constrained model parameters, and two upper limits. It therefore also provides a confirmation that the cluster weak lensing signal indeed scales with source redshift as expected within the employed reference $\Lambda$CDM cosmology.

The HSC data is binned by using weights $W_\varkappa = 1/( \delta g_{t,\varkappa} )^2$, where $\delta g_{t,\varkappa}$ is the estimate for the uncertainty on the tangential reduced shear profile of the $\varkappa$th cluster. The clusters are then binned in redshift bins (0.10, 0.19, 0.27, 0.35, 0.48, 0.8), and count rate bins (0.01, 0.61, 16.1), for which we find a chi-squared of $\chi^{2}_\text{HSC} = 143.4^{+50}_{-16.7}$ with 70 data points, 5 well constrained model parameters. The comparison of the HSC data and the best-fit model is shown in Fig.~\ref{fig:HSC_goodnessoffit}. Some stacked reduced shear profiles clearly deviate from the best-fit model, explaining the high $\chi^{2}$. The fit qualitatively captures the trends in the data.
\section{Summary, discussion, and conclusion}
We present our shear measurements around 237 clusters from the cosmology sample of the first eROSITA All-Sky Survey (eRASS1) in KiDS-1000. We measure the tangential reduced shear in eight geometric annuli ($\SI{250}{\kilo pc}$ wide,  $\SI{0.5}{}-\SI{2.5}{\mega pc}$), separately for each tomographic redshift bin of KiDS. These have been defined for the KiDS cosmic shear analysis, which allows us to directly employ the shear and redshift calibrations derived by the KiDS teams for the KiDS cosmic shear analysis \citep{Hildebrandt2021,van_den_Busch_2022}.

Furthermore we measure and account for cluster member contamination via radial number density profiles of the background galaxies. This analysis includes the injection of simulated galaxy images (using \textit{GalSim}) into the $r$-band mosaics of KiDS-1000, to account for the obscuration caused by foreground and cluster galaxies. We present a global model for both, the radial detection probability and the corrected number density of background-selected sources (which is a measure of the contamination) for eRASS1 clusters in KiDS-1000, based on the cluster redshift and optical richness $\lambda$.

We find that the weighted detection probability typically decreases by 5-12\% at a cluster-centric distance of $\SI{1}{\mega pc}$ due to the obscuration, depending on the tomographic bin as well as the cluster redshift and richness. This would directly translate into identical biases in the number density profiles and boost-corrected reduced shear profiles if unaccounted for (for a more detailed investigation of the effect see \citet{Kleinebreil_2023}. We find that the cluster member contamination has a major impact if the cluster redshift and lower photo-$z$ border of a tomographic bin have a separation of $\Delta z\lesssim 0.2$.\\

The KiDS-1000 weak lensing measurements described in this paper contribute to the joint constraints on cosmological parameters and the eROSITA count rate to halo mass scaling relation presented in \citet{ghirardini23}, together with weak lensing constraints from DES Y3 and HSC-Y3. 
In our paper we present consistency checks for the weak lensing-based constraints obtained from the different data sets.
In particular, we show that differences in the differential surface mass density profiles obtained with different lensing surveys for common galaxy clusters are consistent with zero within the noise as expected. To properly account for the highly correlated intrinsic galaxy shapes for common weak lensing sources between different lensing surveys we introduce bootstrapping conducted on the union of sources. We also compare the scaled Einstein radii estimates on an individual cluster basis, finding no indication for redshift- or richness-dependent biases among the surveys. Furthermore, we compute constraints on the eROSITA count rate to halo mass relation separately for the three weak-lensing surveys. Here we find broadly consistent results, justifying the joint analysis presented in \citet{ghirardini23}. Our analysis contributes to cross-survey comparisons between current stage-III weak-lensing surveys, such as the galaxy-galaxy lensing analysis presented in \citet{Leauthaud_2022}.

Finally, we present stacked tangential reduced shear profiles for eRASS1 clusters both for the KiDS and HSC weak lensing data. Our KiDS analysis conduced in different tomographic redshift bins demonstrates that the cluster weak-lensing signal scales with source redshift as expected in a standard $\Lambda$CDM cosmology, providing a further consistency check for the KiDS weak-lensing measurements. 

Our analysis provides a demonstration for the mass calibration of large galaxy  cluster samples using wide-area weak-lensing surveys. Here our strategy of using the same tomographic source bins employed for cosmic shear analyses provides the major advantage that the shear and redshift calibration can directly be tied to the cosmic shear efforts. This however requires a careful correction for cluster member contamination. For this we suggest that future analyses combine a detection-probability-corrected number-density profile estimation as employed in our work with the photometric redshift  decomposition introduced by \citet{Gruen_2014}, as employed e.g. in the eRASS1 DES analysis by \citet{Grandis_2023}. Combining both approaches  provides an internal consistency check for this important source of systematic uncertainty.
\section*{Acknowledgements}
\tiny
We like to thank Prof. Dr. Henk Hoekstra and Dr. Angus H. Wright for their useful feedback on the manuscript draft.\\

The KiDS-1000 data products are based on observations made with ESO Telescopes at the La Silla Paranal Observatory under programme IDs 177.A-3016, 177.A-3017, 177.A-3018 and 179.A-2004, and on data products produced by the KiDS consortium. The KiDS production team acknowledges support from: Deutsche Forschungsgemeinschaft, ERC, NOVA and NWO-M grants; Target; the University of Padova, and the University Federico II (Naples).\\

F. Kleinebreil, S. Grandis, and T. Schrabback acknowledge support from the German Federal Ministry for Economic Affairs and Energy (BMWi) provided through DLR under projects 50OR2002, 50OR2106, and 50OR2302, support provided by the Deutsche Forschungsgemeinschaft (DFG, German Research Foundation) under grant 415537506, as well as support provided by the Austrian Research Promotion Agency (FFG) and the Federal Ministry of the Republic of Austria for Climate Action, Environment, Mobility, Innovation and Technology (BMK) via the Austrian Space Applications Programme with grant numbers 899537 and 900565.\\

This work is based on data from eROSITA, the soft X-ray instrument aboard SRG, a joint Russian-German science mission supported by the Russian Space Agency (Roskosmos), in the interests of the Russian Academy of Sciences represented by its Space Research Institute (IKI), and the Deutsches Zentrum f{\"{u}}r Luft und Raumfahrt (DLR). The SRG spacecraft was built by Lavochkin Association (NPOL) and its subcontractors and is operated by NPOL with support from the Max Planck Institute for Extraterrestrial Physics (MPE).

The development and construction of the eROSITA X-ray instrument was led by MPE, with contributions from the Dr. Karl Remeis Observatory Bamberg \& ECAP (FAU Erlangen-Nuernberg), the University of Hamburg Observatory, the Leibniz Institute for Astrophysics Potsdam (AIP), and the Institute for Astronomy and Astrophysics of the University of T{\"{u}}bingen, with the support of DLR and the Max Planck Society. The Argelander Institute for Astronomy of the University of Bonn and the Ludwig Maximilians Universit{\"{a}}t Munich also participated in the science preparation for eROSITA.

The eROSITA data shown here were processed using the \esass software system developed by the German eROSITA consortium.
\\

V. Ghirardini, E. Bulbul, A. Liu, C. Garrel, and X. Zhang acknowledge financial support from the European Research Council (ERC) Consolidator Grant under the European Union’s Horizon 2020 research and innovation program (grant agreement CoG DarkQuest No 101002585). N. Clerc was financially supported by CNES. 
%T. Schrabback and F. Kleinebreil acknowledge support from the German Federal
%Ministry for Economic Affairs and Energy (BMWi) provided
%through DLR under projects 50OR2002, 50OR2106, and 50OR2302, as well as the support provided by the Deutsche Forschungsgemeinschaft (DFG, German Research Foundation) under grant 415537506.

The Legacy Surveys consist of three individual and complementary projects: the Dark Energy Camera Legacy Survey (DECaLS; Proposal ID \#2014B-0404; PIs: David Schlegel and Arjun Dey), the Beijing-Arizona Sky Survey (BASS; NOAO Prop. ID \#2015A-0801; PIs: Zhou Xu and Xiaohui Fan), and the Mayall z-band Legacy Survey (MzLS; Prop. ID \#2016A-0453; PI: Arjun Dey). DECaLS, BASS and MzLS together include data obtained, respectively, at the Blanco telescope, Cerro Tololo Inter-American Observatory, NSF’s NOIRLab; the Bok telescope, Steward Observatory, University of Arizona; and the Mayall telescope, Kitt Peak National Observatory, NOIRLab. Pipeline processing and analyses of the data were supported by NOIRLab and the Lawrence Berkeley National Laboratory (LBNL). The Legacy Surveys project is honored to be permitted to conduct astronomical research on Iolkam Du’ag (Kitt Peak), a mountain with particular significance to the Tohono O’odham Nation.
\\

Funding for the DES Projects has been provided by the U.S. Department of Energy, the U.S. National Science Foundation, the Ministry of Science and Education of Spain, the Science and Technology Facilities Council of the United Kingdom, the Higher Education Funding Council for England, the National Center for Supercomputing Applications at the University of Illinois at Urbana-Champaign, the Kavli Institute of Cosmological Physics at the University of Chicago, the Center for Cosmology and Astro-Particle Physics at the Ohio State University, the Mitchell Institute for Fundamental Physics and Astronomy at Texas A\&M University, Financiadora de Estudos e Projetos, Funda{\c c}{\~a}o Carlos Chagas Filho de Amparo {\`a} Pesquisa do Estado do Rio de Janeiro, Conselho Nacional de Desenvolvimento Cient{\'i}fico e Tecnol{\'o}gico and the Minist{\'e}rio da Ci{\^e}ncia, Tecnologia e Inova{\c c}{\~a}o, the Deutsche Forschungsgemeinschaft, and the Collaborating Institutions in the Dark Energy Survey.

The Collaborating Institutions are Argonne National Laboratory, the University of California at Santa Cruz, the University of Cambridge, Centro de Investigaciones Energ{\'e}ticas, Medioambientales y Tecnol{\'o}gicas-Madrid, the University of Chicago, University College London, the DES-Brazil Consortium, the University of Edinburgh, the Eidgen{\"o}ssische Technische Hochschule (ETH) Z{\"u}rich,  Fermi National Accelerator Laboratory, the University of Illinois at Urbana-Champaign, the Institut de Ci{\`e}ncies de l'Espai (IEEC/CSIC), the Institut de F{\'i}sica d'Altes Energies, Lawrence Berkeley National Laboratory, the Ludwig-Maximilians Universit{\"a}t M{\"u}nchen and the associated Excellence Cluster Universe, the University of Michigan, the National Optical Astronomy Observatory, the University of Nottingham, The Ohio State University, the OzDES Membership Consortium, the University of Pennsylvania, the University of Portsmouth, SLAC National Accelerator Laboratory, Stanford University, the University of Sussex, and Texas A\&M University.
\\

The Hyper Suprime-Cam (HSC) collaboration includes the astronomical communities of Japan and Taiwan, and Princeton University. The HSC instrumentation and software were developed by the National Astronomical Observatory of Japan (NAOJ), the Kavli Institute for the Physics and Mathematics of the Universe (Kavli IPMU), the University of Tokyo, the High Energy Accelerator Research Organization (KEK), the Academia Sinica Institute for Astronomy and Astrophysics in Taiwan (ASIAA), and Princeton University. Funding was contributed by the FIRST program from the Japanese Cabinet Office, the Ministry of Education, Culture, Sports, Science and Technology (MEXT), the Japan Society for the Promotion of Science (JSPS), Japan Science and Technology Agency (JST), the Toray Science Foundation, NAOJ, Kavli IPMU, KEK, ASIAA, and Princeton University.

This paper makes use of software developed for Vera C. Rubin Observatory. We thank the Rubin Observatory for making their code available as free software at http://pipelines.lsst.io/.

This paper is based on data collected at the Subaru Telescope and retrieved from the HSC data archive system, which is operated by the Subaru Telescope and Astronomy Data Center (ADC) at NAOJ. Data analysis was in part carried out with the cooperation of Center for Computational Astrophysics (CfCA), NAOJ. We are honored and grateful for the opportunity of observing the Universe from Maunakea, which has the cultural, historical and natural significance in Hawaii.

The Pan-STARRS1 Surveys (PS1) and the PS1 public science archive have been made possible through contributions by the Institute for Astronomy, the University of Hawaii, the Pan-STARRS Project Office, the Max Planck Society and its participating institutes, the Max Planck Institute for Astronomy, Heidelberg, and the Max Planck Institute for Extraterrestrial Physics, Garching, The Johns Hopkins University, Durham University, the University of Edinburgh, the Queen’s University Belfast, the Harvard-Smithsonian Center for Astrophysics, the Las Cumbres Observatory Global Telescope Network Incorporated, the National Central University of Taiwan, the Space Telescope Science Institute, the National Aeronautics and Space Administration under grant No. NNX08AR22G issued through the Planetary Science Division of the NASA Science Mission Directorate, the National Science Foundation grant No. AST-1238877, the University of Maryland, Eotvos Lorand University (ELTE), the Los Alamos National Laboratory, and the Gordon and Betty Moore Foundation.
\\

This work made use of the following Python software packages:
NumPy\footnote{\url{https://numpy.org/}} \citep{Harris_2020},
SciPy\footnote{\url{https://scipy.org/}} \citep{Virtanen_2020}, 
Matplotlib\footnote{\url{https://matplotlib.org/}} \citep{Hunter_2007}, 
Astropy\footnote{\url{https://www.astropy.org/}} \citep{Astropy_2022}.

\bibliographystyle{aa} % style aa.bst
\bibliography{refs.bib}
\normalsize
\appendix

\section{Determining the WL bias}

We shall summarize in the following the basic steps necessary to determine the WL bias and scatter, as well as their uncertainties following the scheme presented in \citet{Grandis_2021}, and its application to eRASS1 clusters described in \citet[][Section 4]{Grandis_2023}. The approach takes the following steps: 
\begin{enumerate}
    \item set up a simulation of realistic reduced shear profiles from surface mass density maps taken from hydro-dynamical simulations, accounting for mis-centering, cluster member contamination and the source redshift distribution and shape measurement biases of the specific survey;
    \item decide on a tangential \tim{reduced} shear profile model to be used in the mass calibration, which will depend on one free mass parameter. We choose a simplified mis-centered NFW \citep{Grandis_2021, bocquet23} with concentration following the relation established by \citet{ragagnin21}, corrected for the mean cluster member contamination;
    \item determine the WL mass such that fitting model fits the synthetic reduced shear profile best. The resulting mass is called weak lensing mass $M_\text{WL}$.
    \item Fit the relation between the halo masses $M$ and the WL masses, \tim{constraining both the} bias and scatter $(b^\text{WL}_l,\,\sigma^\text{WL}_l)$ for each snap shot $l$, as well as a mass trend for both the bias ($b_\text{M}^\text{WL}$), and the scatter ($s_\text{M}$), defined as
\begin{align}
\bigg< \ln \frac{M_{\rm WL}}{M_0} \bigg| M, z_\text{cl} \bigg> =b^\text{WL}_l + b_\text{M}^\text{WL} \ln \left( \frac{M}{M_0} \right)
\nonumber \\
\ln \sigma_{\rm WL}^2 = s_l + s_\text{M} \ln \left( \frac{M}{M_0} \right),
\label{eq:WLbias}
\end{align}
with the pivot mass $M_0 = 2 \times 10^{14}~M_\odot$, and $s_l = \ln  (\sigma^\text{WL}_l)^2$.
\end{enumerate}

We derive the uncertainty on the parameters $(b^\text{WL}_l,\,\sigma^\text{WL}_l,\,b_\text{M}^\text{WL},\,s_\text{M})$ by running Monte Carlo realisations of the simulation prescription above, in which we vary all the input parameters of the hydro-dynamical simulation, that is the parameters of the cluster member contamination, of the mis-centering distribution \citep[described in details in][Section 4.2]{Grandis_2023}, the multiplicative shear bias together with its non-linear part \citep[see][Section 2.1.7]{Grandis_2021}, and the uncertainties on the photometric redshift distribution. We add in quadrature the estimate by \citet[Table 2]{Grandis_2021} on the hydro-dynamical modelling uncertainties, \tim{which currently adds}
%Of most relevance to cosmological applications, is that
a $2\%$ theoretical systematics floor on the WL bias $b_\text{WL}$. 

To compress the information from the Monte Carlo realisations, we perform a principal component analysis on the realisations of $b^\text{WL}_l$, finding that the first two components $\delta_{\text{b}1,2}$ fully describe the scatter around the mean $\mu_\text{b}$. For the variance, one component $\delta_\text{s}$ of scatter around the mean $\mu_\text{s}$ is sufficient. These parameters serve as calibration priors for the halo -- WL scaling relations in the mass calibration analysis.

\subsection{WL mass bias for KiDS-1000}\label{app:bwl}

The determination of the WL bias for KiDS-1000 follows the same synthetic shear model production as described in the context of DES by \citet{Grandis_2023}, Section 4 and summarized above. We however adapt the following points to the KiDS specifications
\begin{itemize}
    \item we use the source redshift distributions from KiDS-1000 (Fig. \ref{fig:tbins}) and vary their mean redshift coherently based on \citet{van_den_Busch_2022} as summarised in Table \ref{tab:vdB},
    \item we apply the multiplicative shear bias from \citet{van_den_Busch_2022} as summarised in Table \ref{tab:vdB}, which we also use for our shear measurement (Sect.\thinspace\ref{sec:gt_meas}), and
    \item we use the cluster member contamination model derived in Sect.\thinspace\ref{sec:cont_model}.
\end{itemize}

\begin{figure}[ht]
    \centering
    \includegraphics[width=\linewidth]{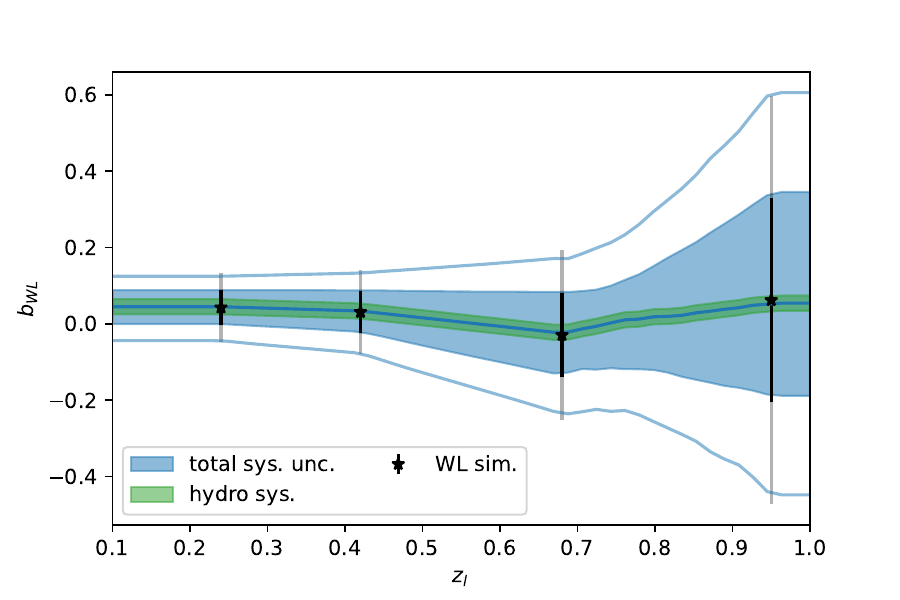}
    \caption{Weak-lensing bias for KiDS-1000 on eRASS1 clusters showing the calibration results on the simulation snapshots as black points and the interpolated trend as blue band ($1\sigma$ uncertainty). At low redshift we are limited by the uncertainty on the multiplicative shear bias, while at higher redshift our uncertainty on the cluster member contamination dominates the systematic error budget ($2\sigma$ uncertainties are indicated by the grey error-bars and light-blue curves). Hydro-dynamical modelling uncertainties, with $1\sigma$ uncertainties shown in green, remain sub-dominant.}
    \label{fig:bWL}
\end{figure}

\begin{table}[ht]
    \centering
    \caption{Multiplicative shear bias $m$ and mean redshift offset $\Delta z$ for the tomographic bins of KiDS-1000 from \citet{van_den_Busch_2022}.}\label{tab:KiDS-multbias}
    \begin{tabular}{ccc}
         photo-$z$ bin & $m$ & $\Delta z=z_\mathrm{est}-z_\mathrm{true}$\\
         \hline
         3 & $-0.011 \pm 0.017$ & $0.013 \pm 0.0116$\\
         4 & $0.008 \pm 0.012$ & $0.011 \pm 0.0084$\\
         5 & $0.012 \pm 0.010$ & $-0.006 \pm 0.0097$\\
    \end{tabular}
    \label{tab:vdB}
\end{table}

\begin{table}
\caption{\label{tab:bWL--kids}
Calibration of the WL bias and scatter for KiDS-1000 WL on eRASS1 clusters.}
\begin{tabular}{lcccc}
$z$ & $0.24$& $0.42$& $0.68$& $0.95$\\
\hline
$\mu_\text{b}$ & $0.043$ & $0.031$ & $-0.029$ & $0.062$ \\  
$\delta_\text{b1}$ & $-0.007$ & $-0.011$ & $-0.013$ & $-0.268$ \\  
$\delta_\text{b2}$ & $0.045$ & $0.056$ & $0.111$ & $0.000$ \\
$\mu_\text{s}$ & $-3.206$ & $-3.217$ & $-2.672$ & $-0.988$ \\  
$\delta_\text{s}$ & $0.303$ & $0.330$ & $0.360$ & $0.315$ \\
\hline
 &  \multicolumn{2}{c}{$b_\text{M}=1.040 \pm 0.026$} &  \multicolumn{2}{c}{$s_\text{M}=-0.638 \pm 0.585$} \\ 
\multicolumn{5}{p{\linewidth-12pt} }{\small Numerical values for the WL bias ($\mu_\text{b}$, $\delta_\text{b1}$, $\delta_\text{b2}$) and scatter ($\mu_\text{s}$, $\delta_{s}$) calibration at different redshifts $z$, and their global mass trends $b_\text{M}$, and $s_\text{M}$, respectively \citep[for details see][]{Grandis_2023}. }
\end{tabular}
\end{table}

We also modify the extraction model, by predicting a model for each tomographic bin
according to its mean
%used by computing each bin's mean 
inverse critical surface mass density 
%(cf. \citet{Grandis_2023}, Eq. 30). 
\citep[compare][Eqn.~30]{Grandis_2023}
The resulting WL bias and scatter numbers are compressed into principal components 
which we report in Table \ref{tab:bWL--kids}. 
We find that at low redshift the uncertainty on the multiplicative shear bias dominates the systematic error budget, 
while the dominating contribution comes from the estimate of the
%while we are limited by the accuracy of our 
cluster member contamination at high redshift.

\subsection{WL mass bias for HSC S19}\label{app:bWL_hsc}

\begin{figure}
  \includegraphics[width=\columnwidth]{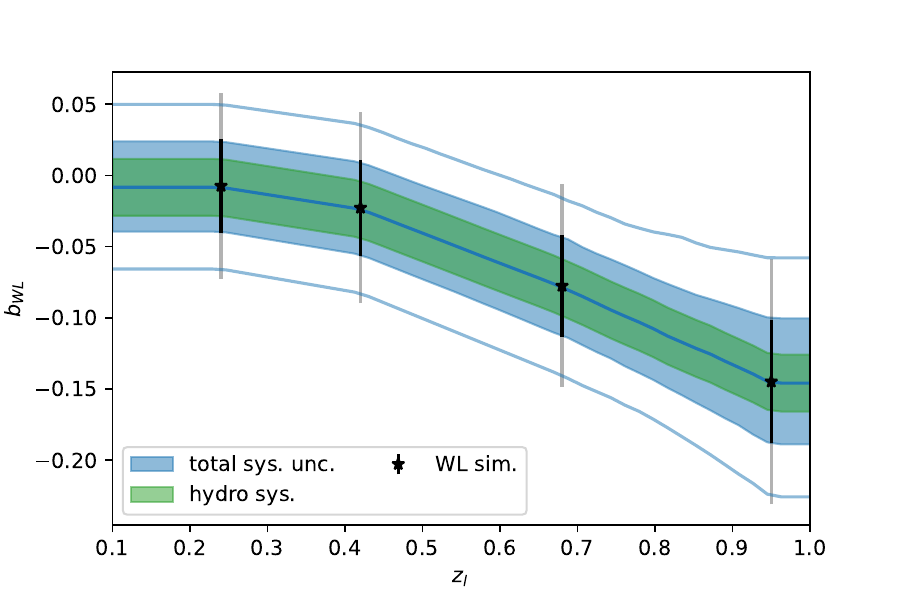}
  \caption{Weak-lensing bias of our extraction model on synthetic shear profile simulations reproducing the HSC WL signal around eRASS1 clusters and groups, showing the calibration results on the simulation snapshots as black points and the interpolated trend as blue band. Contrary to the DES WL bias, the depth of HSC permits a good background selection at all cluster redshifts considered. The larger shape measurement uncertainties however contribute to the WL bias uncertainty even when accounting for the systematics floor from hydro-dynamical modelling uncertainties (green band).}
  \label{fig:hsc-bwl}
\end{figure}

\begin{table}
\caption{\label{tab:bWL--hsc}
Calibration of the WL bias and scatter for HSC WL on eRASS1 clusters.}
\begin{tabular}{lcccc}
$z$ & $0.24$& $0.42$& $0.68$& $0.95$\\  
\hline
$\mu_\text{b}$ & $-0.007$ & $-0.023$ & $-0.078$ & $-0.145$ \\  
$\delta_\text{b1}$ & $-0.030$ & $-0.030$ & $-0.031$ & $-0.038$ \\  
$\delta_\text{b2}$ & $0.007$ & $0.007$ & $0.010$ & $-0.020$ \\  
$\mu_\text{s}$ & $-2.928$ & $-3.142$ & $-2.894$ & $-2.694$ \\  
$\delta_\text{s}$ & $0.267$ & $0.268$ & $0.278$ & $0.283$ \\  
\hline
 &  \multicolumn{2}{c}{$b_\text{M}=1.047 \pm 0.021$} &  \multicolumn{2}{c}{$s_\text{M}=-0.988 \pm 0.601$} \\ 
\multicolumn{5}{p{\linewidth-12pt} }{\small Numerical values for the WL bias ($\mu_\text{b}$, $\delta_\text{b1}$, $\delta_\text{b2}$) and scatter ($\mu_\text{s}$, $\delta_{s}$) calibration at different redshifts $z$, and their global mass trends $b_\text{M}$, and $s_\text{M}$, respectively. }
\end{tabular}
\end{table}

We also perform a weak-lensing bias and scatter  determination for the HSC shear measurements, described in \citet[Section~2.3.1]{ghirardini23}. That analysis follows the same steps as \citet{chiu22}. In our weak lensing bias determination \citep[see][Section~4]{Grandis_2023}  we introduce many modelling differences to the WL bias determination carried out in \citet{chiu22}, Section~4.2. We therefore re-run the weak-lensing bias determination for HSC following the prescriptions made in this work. To account for the weak-lensing specifications of HSC we use
\begin{itemize}
    \item the 
    %synthetic source distributions of the real 
    calibrated source redshift distributions described in \citet[Section~3.8]{chiu22}, 
    %Section~3.8,
    \item the photo-$z$ uncertainty described in the same section,
    \item the same multiplicative shear bias uncertainty as employed in \citet{chiu22}, and
    \item the same upper limit on the cluster member contamination of $<6\%$ that was derived in \citet[Section~3.6]{chiu22}.
    %, Section~3.6.
\end{itemize}

After performing the extraction on many Monte Carlo realisations of the thusly created synthetic shear profiles, we derive the HSC weak-lensing bias shown in Fig.~\ref{fig:hsc-bwl}. Contrary to the DES and KiDS WL bias uncertainty, the clean background selection implemented by \citet{chiu22}, which is enabled by the extraordinary depth of the HSC data, makes the impact of photometric redshift uncertainties negligible. Instead, we find that uncertainties in the calibration of shape measurements contribute significantly to the uncertainty, even when compared to the current systematics floor given by the hydro-dynamical modelling.

\section{Magnification and source number densities}
\label{app:magnification}
Gravitational lenses cause magnification, which affects the number density of WL sources in several ways. To evaluate the impact thereof we closely follow \citet[Section~6.7]{Schrabback_2017}. 

We use COSMOS2020 \texttt{CLASSIC} \citep{Cosmos2020} as a source catalog that is complete up to KiDS-1000's limiting magnitude. For simplification we assume that the COSMOS photo-$z$s roughly match those of KiDS-1000, such that we can divide COSMOS2020 into identical tomographic bins. We then calculate the lensing efficiency $\beta(z_i)$ of each galaxy $i$ in the catalog
\begin{equation}
    \beta(z_i)=\mathrm{max}\left[0,\,\frac{D_\mathrm{ls}}{D_\mathrm{s}}\right],
\end{equation}
where $D_\mathrm{ls}$ and $D_\mathrm{s}$ are the angular diameter distances from the lens to the source and to the source respectively. We choose a representative lens redshift of $z_\mathrm{cl}=0.2$ for our testing. We approximate the magnification that every source experiences as
\begin{equation}
    \mu(z_i)\simeq \frac{\beta(z_i)}{\beta_{0,b}}\left(\mu_{0,b}-1\right)+1,
\end{equation}
where $\mu_{0,b}$ is the magnification at an arbitrary fiducial lensing efficiency $\beta_{0,b}$, for which we use the redshift centre of each tomographic bin (i.e. $\beta_{0,b}(z)=\beta_{0,b}(0.6)$ for the tomographic bin $0.5<z_\mathrm{B}<0.7$).

Magnification can both increase or decrease the source number density. It can increase it as it raises the brightness of sources according to
\begin{equation}
    \mathrm{MAG}_r^\mathrm{lensed}=\mathrm{MAG}_r-2.5\log_{10}\mu(z_i).
\end{equation}
This increase in brightness leads to fainter galaxies being detected that otherwise would not be included in the sample (we use the COSMOS2020 $r$-band magnitudes as the KiDS object detection relies on its respective $r$-band mosaics). On the other hand magnification increases the observed sky area, which leads to a dilution of the source number density 
\tim{by a factor}
%that is equal to 
$1/\mu$, for which we use $\mu_{0,b}$.

\begin{figure}[ht]
    \centering
    \includegraphics[width=\columnwidth]{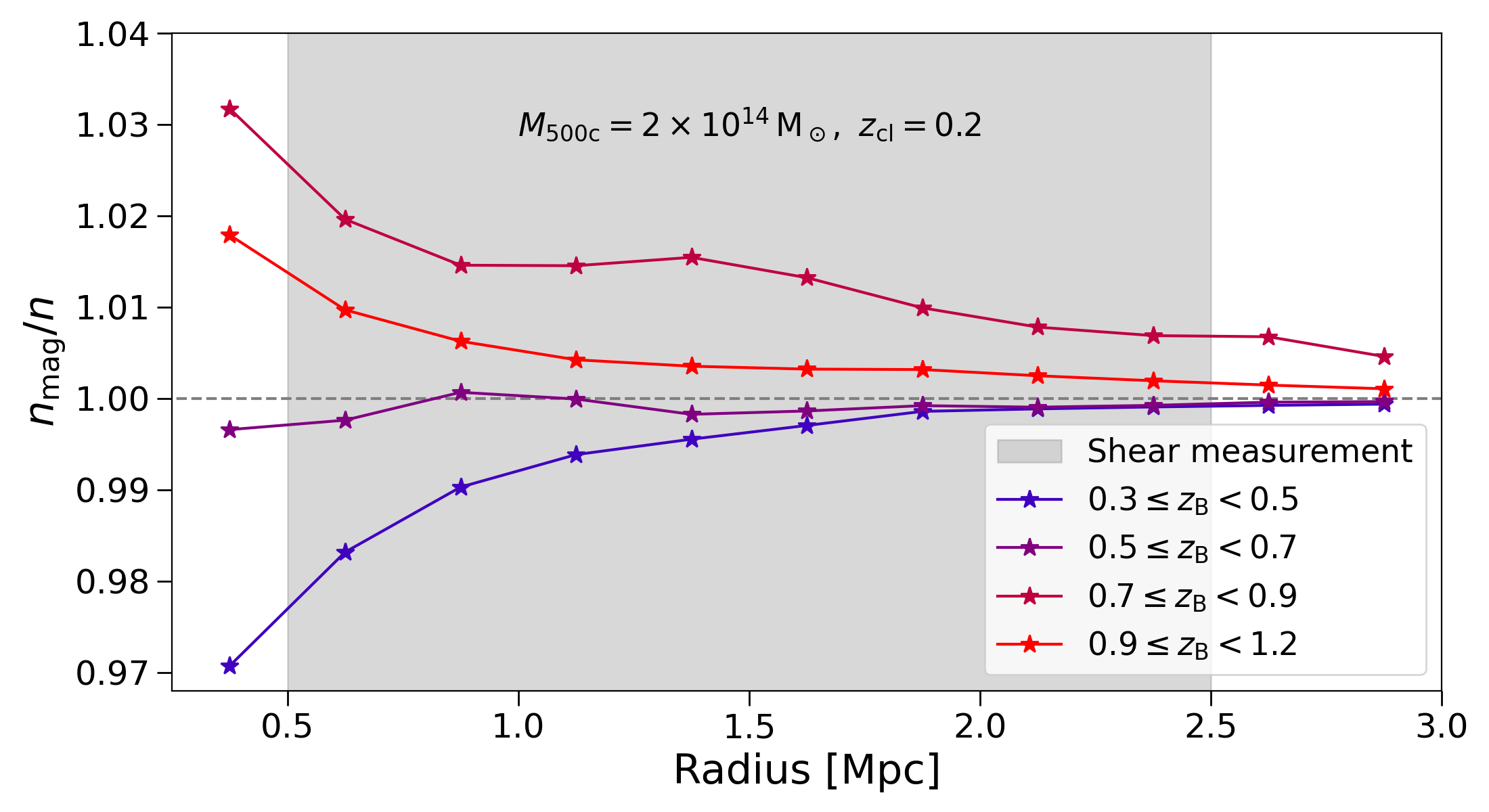}
    \caption{Calculated change in source number density in the different tomographic bins of KiDS-1000 for an NFW halo of mass $M_\mathrm{500c}=\SI{2e14}{M_\odot}$, $z_\mathrm{cl}=0.2$. In the low-redshift tomographic bins the increase in observed sky area (leading to a dilution of the source number density) dominates over the source number density increase due to fainter galaxies which move into the sample, whereas the opposite is the case for the high-redshift tomographic bins.}
    \label{fig:nd_mag}
\end{figure}

To keep track of both effects we calculate the magnification caused by an $M_{500c}=\SI{2e14}{M_\odot}$ NFW halo (at redshift 0.2) over the radial range of our analysis for each tomographic bin as
\begin{equation}
    \mu_{0,b}(R) \simeq 1+2\kappa_{0,b} (R),
\end{equation}
where $\kappa_{0,b} (R)$ is the convergence of the NFW halo with respect to the fiducial lensing efficiency $\beta_{0,b}$ of each tomographic bin. Using these $\mu_{0,b}$ we magnify the COSMOS2020 $r$-band magnitudes, calculate the relative change in galaxy number counts in magnitude bins, and convolve this profile with the tomographic $r$-band magnitude distributions of KiDS-1000. Together with the increase in observed sky area we obtain an approximate change in source number density for KiDS-1000. We show the result in Fig.~\ref{fig:nd_mag}.

We observe that magnification causes a decrease in source density for low-redshift tomographic bins (which are fairly complete in KiDS-1000), as the dilution of the source density dominates over the gain due to the increased source brightnesses. The opposite is the case for the high-redshift tomographic bins, 
\tim{for which KiDS-1000 probes the brighter and steeper part of the galaxy luminosity function}.
%which are less complete in KiDS-1000, and have more faint sources move into the sample. 
Generally the effect is around the one percent level, except for small cluster centric distances, depending on the tomographic bin. 
We note that our approximation does not account for the \tim{additional second-order effect that the increased source sizes may have on the lensing weights.}
%We note that our approximation does not account for the effect that the increased source sizes has, which increases lensing weights if we were to calculate weighted source density profiles, and also modifies the source detection probability.

Together with the fact that magnification has opposite effects in different tomographic bins (which partly cancel from the perspective of a systematic halo mass bias) and that an accurate convergence profile, and therefore halo mass, is necessary to employ a correction, this leads us to dismiss the effect in our analysis. However, we note that one will need to account for magnification in analyses of upcoming stage-IV WL surveys (which aim to reach percent level accuracy), if they rely on source number densities to assess cluster member contamination.

\section{Errata corrige: Tomographic redshift bin definition}

In the final stage of the internal review, we noticed an error in implementing the tomographic bin definition of KiDS-1000. While we selected galaxies within $z_\mathrm{lower}\leq z_\mathrm{B}<z_\mathrm{upper}$, the correct definition is $z_\mathrm{lower}< z_\mathrm{B}\leq z_\mathrm{upper}$. Hence we assigned galaxies with BPZ estimates exactly on the tomographic bin borders to the wrong tomographic bin. We perform several tests to estimate the impact of this error.

\begin{figure}[ht]
    \centering
    \includegraphics[width=\columnwidth]{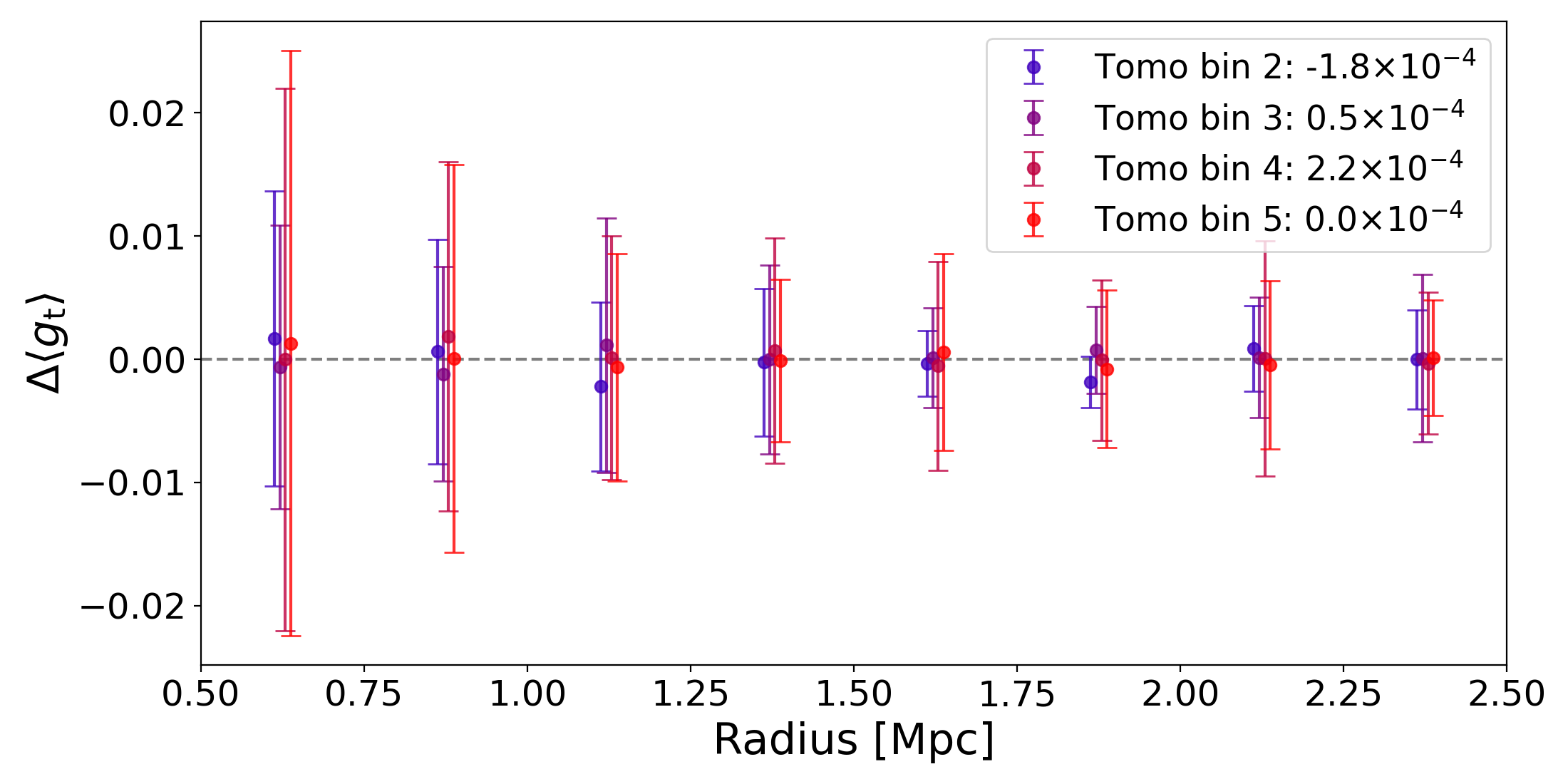}
    \caption{Shift in the stacked weighted reduced tangential shear profiles of the whole cluster sample due to our incorrect tomographic bin definition. The error-bars show uncertainties of the reduced stacked shear profiles themselves, the mean shift per tomographic bin is shown in the legend.}
    \label{fig:gt_diff}
\end{figure}

First, we measure the change in the stacked reduced shear profiles across all lenses, showing the result in Fig. \ref{fig:gt_diff}. We find mean absolute shifts on the order of $10^{-4}$ for the different tomographic bins (which corresponds to relative shifts of 1-2\% as the tangential reduced shear is typically on the order of 1\%), which is well below the statstical uncertainty of our study caused by shape noise. This difference shows that the main conclusions of our work are most likely impervious to this mistake.

\begin{figure}[ht]
    \centering
    \includegraphics[width=\columnwidth]{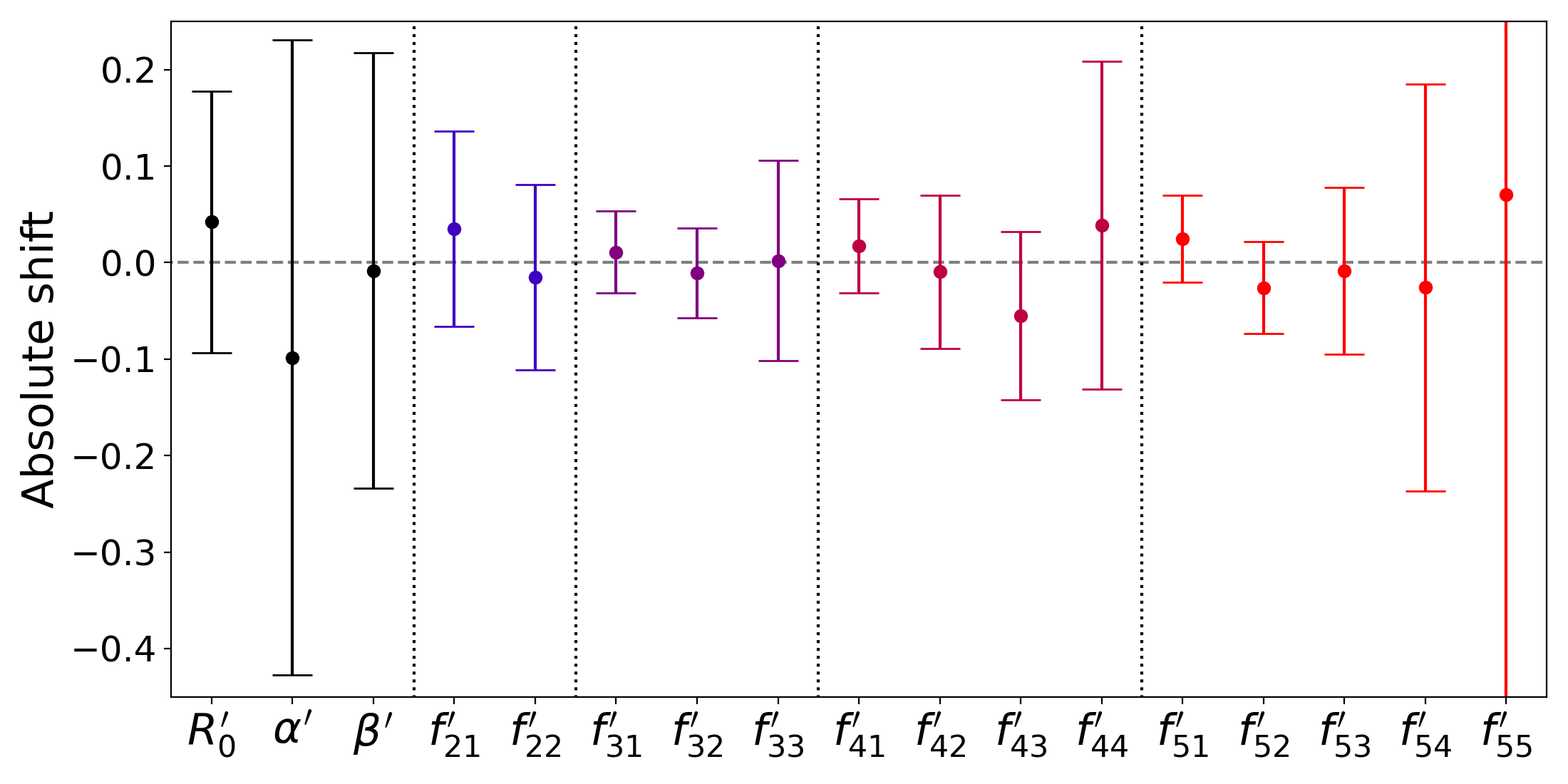}
    \caption{Shift of the contamination model parameters due to our incorrect implementation of the tomographic bin definition. The errors correspond to the uncertainties of the fit parameters. The first index of $f^\prime$ denotes the tomographic bin that a pivot point belongs to, the second index is the position of the pivot point in cluster redshift space.}
    \label{fig:cont_diff}
\end{figure}

Second, we calculate the impact on our contamination model. For this we re-measure the source number density profiles of the whole lens sample with the correct tomographic bin definition, and then use our existing obscuration model to correct them. We argue that the error in the tomographic bin definition is a negligible higher-order effect for the obscuration model, as its amplitude parameters are constant within errors for the different tomographic bins (Fig. \ref{fig:d_zcl}). We extract a new contamination model from these source density profiles, and show the shift of the 17 fit parameters in Fig. \ref{fig:cont_diff}. The mean shift of the contamination model parameters is $0.26\,\sigma$ (ignoring the direction of the shifts although some fit parameters are degenerate and shift in opposite directions). This represents a statistically insignificant impact.

Third, we evaluate the chi-squared of the previous KiDS alone best-fit model with the correctly measured tangential reduced shear profiles. This leads to a new chi-squared $\chi^{2}_\text{KiDS} = 44.5^{+9.8}_{-0.5}$, which constitutes a statistically insignificant decrease by $\Delta\chi^{2}_\text{KiDS} = 1.5$. When evaluating the impact of the corrected data vector on the best-fit point of the entire cosmological analysis presented in \citet{ghirardini23}, we find a reduction of the chi-squared of the KiDS stacked WL by $\Delta\chi^{2}_\text{KiDS} = 3.3$. These changes are well within the systematic scatter derived from the propagation of our posterior on the scaling relation parameters. 

Given this error's minimal impact, we document it here transparently, but conclude that a full reprocessing with a corrected selection is not required, especially for the main cosmological analysis \citep{ghirardini23}.

\end{document}